\documentclass[a4paper,11pt]{article}
\pdfoutput=1
\usepackage{jheppub}
\usepackage[utf8]{inputenc}
\usepackage{url}
\usepackage{mathrsfs}  
\usepackage{pdflscape}
\usepackage{refcount}
\usepackage{booktabs}
\usepackage{todonotes}
\usepackage{enumitem}
\usepackage{adjustbox}
\usepackage{afterpage}

\usepackage{caption}
\usepackage{subcaption}

\usepackage{tikz}
\usetikzlibrary{positioning}
\usetikzlibrary{arrows}
\usetikzlibrary{decorations.pathreplacing}
\usetikzlibrary{shapes.geometric}
\usetikzlibrary{calc}
\usetikzlibrary{decorations.pathmorphing}

\definecolor{goodgreen}{RGB}{55,169,49}

\newcommand{\convexpath}[2]{
  [   
  create hullcoords/.code={
    \global\edef\namelist{#1}
    \foreach [count=\counter] \nodename in \namelist {
      \global\edef\numberofnodes{\counter}
      \coordinate (hullcoord\counter) at (\nodename);
    }
    \coordinate (hullcoord0) at (hullcoord\numberofnodes);
    \pgfmathtruncatemacro\lastnumber{\numberofnodes+1}
    \coordinate (hullcoord\lastnumber) at (hullcoord1);
  },
  create hullcoords
  ]
  ($(hullcoord1)!#2!-90:(hullcoord0)$)
  \foreach [
  evaluate=\currentnode as \previousnode using \currentnode-1,
  evaluate=\currentnode as \nextnode using \currentnode+1
  ] \currentnode in {1,...,\numberofnodes} {
    let \p1 = ($(hullcoord\currentnode) - (hullcoord\previousnode)$),
    \n1 = {atan2(\y1,\x1) + 90},
    \p2 = ($(hullcoord\nextnode) - (hullcoord\currentnode)$),
    \n2 = {atan2(\y2,\x2) + 90},
    \n{delta} = {Mod(\n2-\n1,360) - 360}
    in 
    {arc [start angle=\n1, delta angle=\n{delta}, radius=#2]}
    -- ($(hullcoord\nextnode)!#2!-90:(hullcoord\currentnode)$) 
  }
}

\tikzset{flavour/.style={draw=none,minimum size=0.3mm,fill=white, regular polygon,regular polygon sides=4,draw}}
\tikzset{gaugeBig/.style={inner sep=1mm,draw=none,fill=white,minimum size=2mm,circle, draw}}
\tikzset{bd/.style={circle, draw=black, inner sep=0pt, fill=black, minimum size=2mm}}
\tikzset{wd/.style={circle, draw=black, inner sep=0pt, fill=white, minimum size=2mm}}
\tikzset{Dynkin/.style={circle, draw=black, inner sep=0pt, fill=white, minimum size=2mm}}
\tikzstyle{ligne}=[draw, very thick] 
\tikzstyle{gridline}=[draw, gray] 
\tikzset{gauge/.style={circle, draw,inner sep=2.5pt}}
\tikzset{gaugeo/.style={circle, draw,inner sep=2.5pt,fill=orange}}
\tikzset{gauger/.style={circle, draw,inner sep=2.5pt,fill=red}}
\tikzset{gaugeb/.style={circle, draw,inner sep=2.5pt,fill=blue}}
\tikzset{gaugeg/.style={circle, draw,inner sep=2.5pt,fill=green}}
\tikzset{gaugegoodgreen/.style={circle, draw,inner sep=2.5pt,fill=goodgreen}}
\tikzset{gaugem/.style={circle, draw,inner sep=2.5pt,fill=magenta}}
\tikzset{hasse/.style={circle, fill,inner sep=2pt}}
\tikzset{d2/.style={circle, fill,inner sep=1.3pt}}
\tikzset{shrinky/.style={circle, fill,inner sep=1pt}}
\tikzset{sized/.style={circle, draw, inner sep=1.5pt}}
\tikzset{seven/.style={circle, draw,inner sep=3pt}}

\newcommand{\rect}[5]{
\fill[black!#5] (#1,#2)--(#3,#2)--(#3,#4)--(#1,#4)--(#1,#2);}

\preprint{Imperial/TP/22/AH/01}
\title{The Hasse Diagram of the Moduli Space of Instantons}
\author[/ \hspace*{-0.048cm}\backslash]{Antoine Bourget,}
\author[|]{Julius F. Grimminger,}
\author[|]{Amihay Hanany,}
\author[|]{and Zhenghao Zhong\,}
\affiliation[/]{Université Paris-Saclay, CNRS, CEA, Institut de physique théorique, 91191, Gif-sur-Yvette, France}
\affiliation[\backslash]{Laboratoire de Physique de l'\'Ecole Normale Sup\'erieure, PSL University, \\ 24 rue Lhomond, 75005 Paris, France}
\affiliation[|]{Theoretical Physics Group, The Blackett Laboratory, Imperial College London, Prince Consort Road
London, SW7 2AZ, UK}
\emailAdd{antoine.bourget@polytechnique.org}
\emailAdd{julius.grimminger17@imperial.ac.uk}
\emailAdd{a.hanany@imperial.ac.uk}
\emailAdd{zhenghao.zhong14@imperial.ac.uk}

\abstract{Hasse diagrams (or phase diagrams) for moduli spaces of supersymmetric field theories have been intensively studied in recent years, and many tools to compute them have been developed. The moduli space of instantons, despite being well studied, has proven difficult to deal with. In this note we explore the Hasse diagram of this moduli space from several perspectives -- using the partial Higgs mechanism, using brane systems and using quiver subtraction -- having to refine previously developed techniques. In particular we introduce the new concept of \emph{decorated quiver}, which allows to deal with a large class of unitary quivers, including those with adjoint matter.  }

\begin{document}
\maketitle

\section{Introduction}

Moduli spaces of instantons, despite being well studied objects, are notoriously difficult to understand. These moduli spaces are examples of symplectic singularities, which have recently received a lot of attention from both mathematicians and physicists. Since the work of Kraft and Procesi Hasse diagrams are used to understand properties of symplectic singularities. If the symplectic singularity in question is a slice in the nilpotent cone of a finite dimensional Lie algebra, then its Hasse diagram is obtained from that of the nilpotent cone. For classical algebras this was computed by Kraft and Procesi \cite{kraft1980minimal,Kraft1982}. For the exceptional groups this was done more recently \cite{2015arXiv150205770F}. On the other hand, if the symplectic singularity in question is the ($3d$ $\mathcal{N}=4$) Coulomb branch of a quiver, then, under certain conditions of the quiver being `nice enough', one can use quiver subtraction\footnote{If a brane system for the quiver exists, quiver subtractions are equivalent to Kraft-Procesi transitions \cite{Cabrera:2016vvv}, see e.g. \cite[Sec.\ 2.6]{Gaiotto:2013bwa} and \cite{Cabrera:2017njm,Bourget:2019aer}.} \cite{Cabrera:2018ann,Bourget:2019aer} to compute the Hasse diagram. It turns out that the quivers whose Coulomb branches are moduli spaces of instantons are not nice enough to be tackled with conventional quiver subtraction. The problem is essentially that the same slice can be subtracted multiple times in a row on the same set of nodes.\footnote{This places the moduli space of instantons in the realm of the double affine Grassmannian, as opposed to the standard affine Grassmannian. The standard affine Grassmannian is well understood through conventional quiver subtraction \cite{Bourget:2021siw}.} This problem is partially solved in \cite[Appendix C]{Bourget:2020mez} by introducing a \emph{decoration} for quivers as a bookkeeping tool in the quiver subtraction. It turns out that in order to understand all elementary transitions in the moduli space of instantons these decorated quivers need to be interpreted as theories in their own right (or at least as a definition of a Coulomb branch), much like the non-simply laced quivers. In this work we focus solely on instantons on $\mathbb{R}^4$, and not on any of its quotients.

\begin{figure}
    \centering
\scalebox{.8}{\begin{tikzpicture}
\rect{1.5}{1.5}{-16.5}{-14.5}{5};
\rect{1.5}{1.5}{-13.5}{-11.5}{10};
\rect{1.5}{1.5}{-10.5}{-8.5}{15};
\rect{1.5}{1.5}{-7.5}{-5.5}{20};
\rect{1.5}{1.5}{-4.5}{-2.5}{25};
                \node[hasse] (1) at (0,0) {};
                \node[hasse] (2) at (-3,-1) {};
                \node[hasse] (3) at (-6,-2) {};
                \node[hasse] (4) at (-6,-4) {};
                \node[hasse] (5) at (-9,-3) {};
                \node[hasse] (6) at (-9,-5) {};
                \node[hasse] (7) at (-9,-7) {};
                \node[hasse] (8) at (-12,-4) {};
                \node[hasse] (9) at (-12,-6) {};
                \node[hasse] (10) at (-12.5,-8) {};
                \node[hasse] (11) at (-11.5,-8) {};
                \node[hasse] (12) at (-12,-10) {};
                \node[hasse] (13) at (-15,-5) {};
                \node[hasse] (14) at (-15,-7) {};
                \node[hasse] (15) at (-15.5,-9) {};
                \node[hasse] (16) at (-14.6,-9) {};
                \node[hasse] (17) at (-16,-11) {};
                \node[hasse] (18) at (-15,-11) {};
                \node[hasse] (19) at (-15.5,-13) {};
                \draw[red] (1)--(2)--(3)--(5)--(8)--(13) (4)--(6)--(9)--(14) (7)--(11)--(16) (10)--(15) (12)--(18);
                \draw[dashed] (6)--(7) (9)--(11)--(12) (14)--(16)--(18)--(19)--(17)--(15);
                \draw (3)--(4) (5)--(6) (8)--(9)--(10)--(12) (13)--(14)--(15)--(18) (16)--(17);
                \node at (-3,-2) {$\mathcal{M}_{1,G}$};
                \node at (-6,-5) {$\mathcal{M}_{2,G}$};
                \node at (-9,-8) {$\mathcal{M}_{3,G}$};
                \node at (-12,-11) {$\mathcal{M}_{4,G}$};
                \node at (-15.5,-14) {$\mathcal{M}_{5,G}$};
\node at (-3,-13) {\begin{tikzpicture}
                \draw[red] (-4,-12.3)--(-3,-12.3);
                \draw[black] (-4,-13)--(-3,-13);
                \draw[dashed] (-4,-13.7)--(-3,-13.7);
                \node at (-1,-12.3) {$\overline{\mathcal{O}}_{\textrm{min}} (\mathfrak{g})$};
                \node at (-1,-13) {$A_1$};
                \node at (-1,-13.7) {$m$};
                \node at (1,-11.6) {$\mathrm{dim}_{\mathbb{H}}$};
                \node at (1,-12.3) {$h^{\vee}-1$};
                \node at (1,-13) {$1$};
                \node at (1,-13.7) {$1$};
        \end{tikzpicture}};
        \end{tikzpicture}
}
    \caption{Hasse diagrams for the moduli spaces $\mathcal{M}_{k,G}$ of $k$ $G$-instantons on $\mathbb{R}^4$ for $1 \leq k \leq 5$. Only three types of elementary slices show up: the closure of the minimal orbit of the Lie algebra of $G$, the Kleinian singularity $A_1 := \mathbb{C}^2 / \mathbb{Z}_2$ and the non-normal $m$ (see Appendix \ref{app:G2} for the definition of $m$). The quaternionic dimension of the transverse slice from the bottom to the top leaf of $\mathcal{M}_{k,G}$ is $k \cdot (h^{\vee}-1) + (k-1) \cdot 1 = k h^{\vee} -1$ .   }
    \label{fig:summary}
\end{figure}
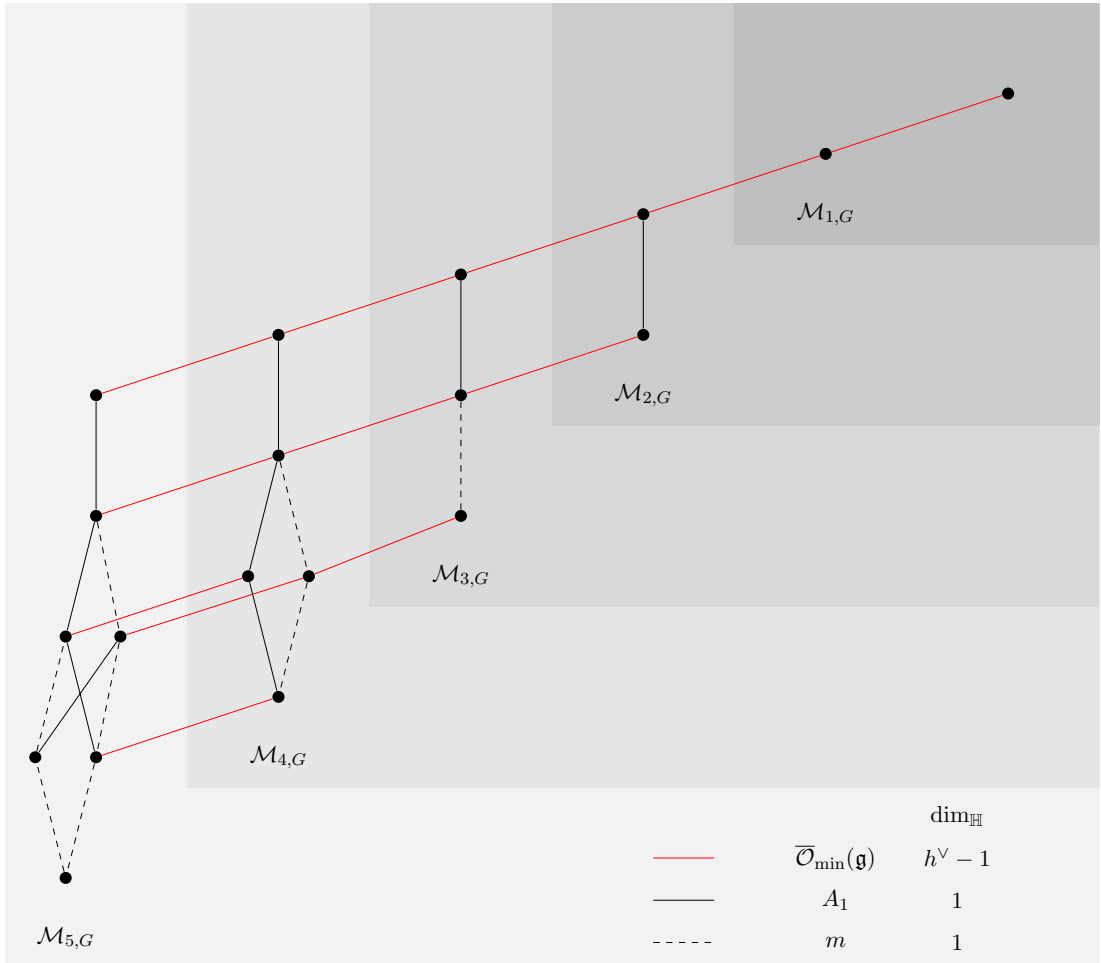

Our results are summarised in Figure \ref{fig:summary}. This paper is dedicated to establishing the Hasse diagrams in that figure using various methods.

\paragraph{Organisation of the paper} In Section \ref{sec:generalities} we recall basic facts about moduli spaces of instantons. This section may be skipped by readers familiar with instantons and D-branes. In Section \ref{sec:main} we use every tool at our disposal to study the Hasse diagram of ththese moduli spaces, including the Higgs mechanism, brane systems, and quiver subtraction. In Section \ref{sec:Conclusion} we summarise our findings, and discuss open problems. In Appendix \ref{app:G2} we discuss various slices in the nilcone of $G_2$ and comment on the slice $m$. In Appendix \ref{app:SYM} we discuss some symmetric products of $\mathbb{C}^2$ and compute their Hasse diagrams. In Appendix \ref{app:braneweb} we discuss refined rules for reading magnetic quivers from brane web decompositions, allowing for decorated magnetic quivers and short nodes.

\section{Basic facts on the moduli space of instantons}
\label{sec:generalities}

In the following we refer to the moduli space of $k$ $G$-instantons on $\mathbb{R}^4$ as $\mathcal{M}_{k,G}$. The dimension of this moduli space is the product of the instanton number $k$ and the dual coxeter number $h^\vee_G$ of $G$,
\begin{equation}
    \mathrm{dim}_\mathbb{H}\left(\mathcal{M}_{k,G}\right)=kh^\vee_G\,.
\end{equation}
This moduli space contains a free factor, $\mathbb{H}$, which corresponds to the overall position. The global symmetry algebra of $\mathcal{M}_{k,G}$ is
\begin{equation}
    \begin{tabular}{cc}
        $k=1:$ & $A_1\mathfrak{g}$ \\
        $k>1:$ & $A_1A_1\mathfrak{g}$
    \end{tabular}
\end{equation}
where one $A_1$ is the global symmetry of the free part, and $\mathfrak{g}=\mathrm{Lie}(G)$.

The moduli space of instantons, while simple to define, is a complicated object to describe. Several advances were made to make this object more tangible:

\paragraph{The ADHM construction \cite{atiyah1994construction}} of $\mathcal{M}_{k,G}$, for classical groups $G$, is a realisation as a hyper-K\"ahler quotient. This allows to identify $\mathcal{M}_{k,G}$ with the Higgs branch of $4$d $\mathcal{N}=2$ (or $3$d $\mathcal{N}=4$,\ etc.) SQCD theories with two types of matter \cite{Witten:1995gx}. The relevant ADHM-quivers are summarised in Figure \ref{fig:ADHM}.

\paragraph{Branes within Branes \cite{Witten:1995gx,Douglas:1995bn}} realises $\mathcal{M}_{k,G}$ as the moduli space of $k$ D$p$ branes inside a stack of coincident $(p+4)$-branes. Classical $G$ can be realised as a gauge group on a stack of D($p+4$) branes, possibly in the presence of an O$(p+4)$ orientifold plane, given that the orientifold plane exists. For $p=3$ every $G=ADE$ can be realised as the gauge group on a stack of $7$-branes. The worldvolume theory of the D$p$ branes is that of an ADHM quiver, or a rank-$k$ $E_N$ theory \cite{Dasgupta:1996ij,Minahan:1996fg,Minahan:1996cj}. The Coulomb branch of this theory can be visualised as the moduli of the D$p$ branes moving transverse to the $(p+4)$-branes. The Higgs branch of this theory, i.e.\ our moduli space of instantons, is more involved. 
Any single D$p$ brane may bind together with the stack of $(p+4)$-branes, in the sense that it loses its interpretation as a brane and becomes a (self-dual) gauge field on the worldvolume theory on the $(p+4)$-branes. 
There is an asymptotic region in the moduli space where the $k$ instantons can be considered independently, thereby singling out in that region $k$ quaternionic position moduli. By contrast, at a generic point in the moduli space this interpretation is lost. However tuning appropriate $h^\vee_G-1$ moduli realises a small instanton transition. After this transition, the D$p$ brane interpretation is recovered and it carries one position modulus.

\begin{figure}
     \centering
     \begin{subfigure}[b]{0.2\textwidth}
         \centering
             \begin{tikzpicture}
                \node[flavour,label=below:{$A_{N-1}$}] (f) at (0,0) {};
                \node[gaugeBig, label=below:{$\mathrm{U}(k)$}] (g) at (1,0) {};
                \draw (f)--(g);
                \draw (g) to [out=315,in=45,looseness=10] (g);
                \node at (1.9,0) {Adj};
            \end{tikzpicture}
         \caption{}
         \label{fig:ADHM_A}
     \end{subfigure}
     \begin{subfigure}[b]{0.2\textwidth}
         \centering
             \begin{tikzpicture}
                \node[flavour,label=below:{$B_N$}] (f) at (0,0) {};
                \node[gaugeBig, label=below:{Sp${}^\prime(k)$}] (g) at (1,0) {};
                \draw (f)--(g);
                \draw (g) to [out=315,in=45,looseness=10] (g);
                \node at (1.75,0) {A};
            \end{tikzpicture}
         \caption{}
         \label{fig:ADHM_B}
     \end{subfigure}
     \begin{subfigure}[b]{0.2\textwidth}
         \centering
             \begin{tikzpicture}
                \node[flavour,label=below:{$C_N$}] (f) at (0,0) {};
                \node[gaugeBig, label=below:{O$(k)$}] (g) at (1,0) {};
                \draw (f)--(g);
                \draw (g) to [out=315,in=45,looseness=10] (g);
                \node at (1.75,0) {S};
            \end{tikzpicture}
         \caption{}
         \label{fig:ADHM_C}
     \end{subfigure}
     \begin{subfigure}[b]{0.2\textwidth}
         \centering
             \begin{tikzpicture}
                \node[flavour,label=below:{$D_N$}] (f) at (0,0) {};
                \node[gaugeBig, label=below:{Sp$(k)$}] (g) at (1,0) {};
                \draw (f)--(g);
                \draw (g) to [out=315,in=45,looseness=10] (g);
                \node at (1.75,0) {A};
            \end{tikzpicture}
         \caption{}
         \label{fig:ADHM_D}
     \end{subfigure}
        \caption{ADHM quivers whose Higgs branch is $\mathcal{M}_{k,G}$. $A$ denotes the 2nd antisymmetric, and $S$ the 2nd symmetric of the fundamental representation (these are not irreducible as they include singlets). We have: (a) $G=\mathrm{SU}(N)$, (b) $G=\mathrm{SO}(2N+1)$, (c) $G=\mathrm{Sp}(N)$, (d) $G=\mathrm{SO}(2N)$. It is essential that the gauge group in (c) is $O(k)$ and not $SO(k)$. }
        \label{fig:ADHM}
\end{figure}

\paragraph{The CFHM construction \cite{Cremonesi:2014xha}} of $\mathcal{M}_{k,G}$ allows to construct it as a moduli space of dressed monopole operators -- more precisely the Coulomb branch of a $3$d $\mathcal{N}=4$ theory \eqref{eq:CFHM}. This is possible for both classical and exceptional algebras. For classical algebras the CFHM quivers were derived from T-dualising the branes within branes setups of the ADHM quivers. The quivers whose Coulomb branch is $\mathcal{M}_{k,G}$ have the general form
\begin{equation}
    \mathsf{Q}_{k,G}=
\vcenter{\hbox{\scalebox{1}{
    \begin{tikzpicture}
    \node[gauge,label=below:{$1$}] (1) at (0,0) {};
    \node[gauge] (2) at (1,0) {$\hat{\mathsf{Q}}_{k,G}$};
    \draw (1)--(2);
    \end{tikzpicture}}}}\,,
    \label{eq:CFHM}
\end{equation}
where $\hat{\mathsf{Q}}_{k,G}$ denotes the Dynkin quiver of affine $G$ with all nodes multiplied by $k$. An extra $\mathrm{U}(1)$ is connected to the affine node to make $\mathsf{Q}_{k,G}$. We display the CFHM quivers in Figure \ref{fig:CFHMquivers}. Some comments are in order:
\begin{enumerate}
    \item For $k>1$ the quiver $\hat{\mathsf{Q}}_{k,G}$ is bad in the classification of \cite{Gaiotto:2008ak}, even though all of its nodes are balanced. $\mathsf{Q}_{k,G}$, on the other hand, is always good.
    \item For $G$ classical $\mathsf{Q}_{k,G}$ are 3d mirror duals of the ADHM quivers.
\end{enumerate}

\begin{figure}
    \centering
    \begin{tikzpicture}
        \node[label=left:{(a)}] (a) at (0,0) {$
            \begin{tikzpicture}
                \node[gauge,label=below:{$1$}] (0) at (0,0) {};
                \node[gauge,label=below:{$k$}] (1) at (1,0) {};
                \node (2) at (2,0) {$\cdots$};
                \node[gauge,label=below:{$k$}] (3) at (3,0) {};
                \node[gauge,label=left:{$k$}] (u) at (2,1) {};
                \draw (0)--(1)--(2)--(3)--(u)--(1);
                \draw [decorate,decoration={brace,amplitude=5pt}] (3.2,-0.6)--(0.8,-0.6);
                \node at (2,-1) {$N-1$};
            \end{tikzpicture}
        $};
        \node[label=left:{(b)}] (b) at (8,0) {$
            \begin{tikzpicture}
                \node[gauge,label=below:{$1$}] (0) at (0,0) {};
                \node[gauge,label=below:{$k$}] (1) at (1,0) {};
                \node[gauge,label=left:{$k$}] (2u) at (2,1) {};
                \node[gauge,label=below:{$2k$}] (2) at (2,0) {};
                \node (3) at (3,0) {$\cdots$};
                \node[gauge,label=below:{$2k$}] (4) at (4,0) {};
                \node[gauge,label=below:{$k$}] (5) at (5,0) {};
                \draw (0)--(1)--(2)--(2u) (2)--(3)--(4);
                \draw[transform canvas={yshift=1pt}] (4)--(5);
                \draw[transform canvas={yshift=-1pt}] (4)--(5);
                \draw (4.5-0.1,0.2)--(4.5+0.1,0)--(4.5-0.1,-0.2);
                \draw [decorate,decoration={brace,amplitude=5pt}] (5.2,-0.6)--(0.8,-0.6);
                \node at (3,-1) {$N$};
            \end{tikzpicture}
        $};
        \node[label=left:{(c)}] (c) at (0,-3) {$
            \begin{tikzpicture}
                \node[gauge,label=below:{$1$}] (0) at (0,0) {};
                \node[gauge,label=below:{$k$}] (1) at (1,0) {};
                \node[gauge,label=below:{$k$}] (2) at (2,0) {};
                \node (3) at (3,0) {$\cdots$};
                \node[gauge,label=below:{$k$}] (4) at (4,0) {};
                \node[gauge,label=below:{$k$}] (5) at (5,0) {};
                \draw (0)--(1) (2)--(3)--(4);
                \draw[transform canvas={yshift=1pt}] (1)--(2);
                \draw[transform canvas={yshift=-1pt}] (1)--(2);
                \draw (1.5-0.1,0.2)--(1.5+0.1,0)--(1.5-0.1,-0.2);
                \draw[transform canvas={yshift=1pt}] (4)--(5);
                \draw[transform canvas={yshift=-1pt}] (4)--(5);
                \draw (4.5+0.1,0.2)--(4.5-0.1,0)--(4.5+0.1,-0.2);
                \draw [decorate,decoration={brace,amplitude=5pt}] (5.2,-0.6)--(1.8,-0.6);
                \node at (3.5,-1) {$N$};
            \end{tikzpicture}
        $};
        \node[label=left:{(d)}] (d) at (8,-3) {$
            \begin{tikzpicture}
                \node[gauge,label=below:{$1$}] (0) at (0,0) {};
                \node[gauge,label=below:{$k$}] (1) at (1,0) {};
                \node[gauge,label=left:{$k$}] (2u) at (2,1) {};
                \node[gauge,label=below:{$2k$}] (2) at (2,0) {};
                \node (3) at (3,0) {$\cdots$};
                \node[gauge,label=left:{$k$}] (4u) at (4,1) {};
                \node[gauge,label=below:{$2k$}] (4) at (4,0) {};
                \node[gauge,label=below:{$k$}] (5) at (5,0) {};
                \draw (0)--(1)--(2)--(2u) (2)--(3)--(4)--(4u) (4)--(5);
                \draw [decorate,decoration={brace,amplitude=5pt}] (5.2,-0.6)--(0.8,-0.6);
                \node at (3,-1) {$N-1$};
            \end{tikzpicture}
        $};
        \node[label=left:{(e6)}] (e6) at (0,-6) {$
            \begin{tikzpicture}
                \node[gauge,label=below:{$1$}] (0) at (0,0) {};
                \node[gauge,label=below:{$k$}] (1) at (1,0) {};
                \node[gauge,label=below:{$2k$}] (2) at (2,0) {};
                \node[gauge,label=below:{$3k$}] (3) at (3,0) {};
                \node[gauge,label=left:{$2k$}] (3u) at (3,1) {};
                \node[gauge,label=left:{$k$}] (3uu) at (3,2) {};
                \node[gauge,label=below:{$2k$}] (4) at (4,0) {};
                \node[gauge,label=below:{$k$}] (5) at (5,0) {};
                \draw (0)--(1)--(2)--(3)--(3u)--(3uu) (3)--(4)--(5);
            \end{tikzpicture}
        $};
        \node[label=left:{(e7)}] (e7) at (8,-6) {$
            \begin{tikzpicture}
                \node[gauge,label=below:{$1$}] (0) at (0,0) {};
                \node[gauge,label=below:{$k$}] (1) at (1,0) {};
                \node[gauge,label=below:{$2k$}] (2) at (2,0) {};
                \node[gauge,label=below:{$3k$}] (3) at (3,0) {};
                \node[gauge,label=below:{$4k$}] (4) at (4,0) {};
                \node[gauge,label=left:{$2k$}] (4u) at (4,1) {};
                \node[gauge,label=below:{$3k$}] (5) at (5,0) {};
                \node[gauge,label=below:{$2k$}] (6) at (6,0) {};
                \node[gauge,label=below:{$k$}] (7) at (7,0) {};
                \draw (0)--(1)--(2)--(3)--(4)--(4u) (4)--(5)--(6)--(7);
            \end{tikzpicture}
        $};
        \node[label=left:{(e8)}] (e8) at (4,-8.5) {$
            \begin{tikzpicture}
                \node[gauge,label=below:{$1$}] (0) at (0,0) {};
                \node[gauge,label=below:{$k$}] (1) at (1,0) {};
                \node[gauge,label=below:{$2k$}] (2) at (2,0) {};
                \node[gauge,label=below:{$3k$}] (3) at (3,0) {};
                \node[gauge,label=below:{$4k$}] (4) at (4,0) {};
                \node[gauge,label=below:{$5k$}] (5) at (5,0) {};
                \node[gauge,label=below:{$6k$}] (6) at (6,0) {};
                \node[gauge,label=left:{$3k$}] (6u) at (6,1) {};
                \node[gauge,label=below:{$4k$}] (7) at (7,0) {};
                \node[gauge,label=below:{$2k$}] (8) at (8,0) {};
                \draw (0)--(1)--(2)--(3)--(4)--(5)--(6)--(6u) (6)--(7)--(8);
            \end{tikzpicture}
        $};
        \node[label=left:{(f4)}] (f4) at (0,-10.5) {$
            \begin{tikzpicture}
                \node[gauge,label=below:{$1$}] (m2) at (-2,0) {};
                \node[gauge,label=below:{$k$}] (m1) at (-1,0) {};
                \node[gauge,label=below:{$2k$}] (0) at (0,0) {};
                \node[gauge,label=below:{$3k$}] (1) at (1,0) {};
                \node[gauge,label=below:{$2k$}] (2) at (2,0) {};
                \node[gauge,label=below:{$k$}] (3) at (3,0) {};
                \draw (m2)--(m1)--(0)--(1) (2)--(3);
                \draw[transform canvas={yshift=1pt}] (1)--(2);
                \draw[transform canvas={yshift=-1pt}] (1)--(2);
                \draw (1.5-0.1,0.2)--(1.5+0.1,0)--(1.5-0.1,-0.2);
            \end{tikzpicture}
        $};
        \node[label=left:{(g2)}] (g2) at (8,-10.5) {$
            \begin{tikzpicture}
                \node[gauge,label=below:{$1$}] (m1) at (-1,0) {};
                \node[gauge,label=below:{$k$}] (0) at (0,0) {};
                \node[gauge,label=below:{$2k$}] (1) at (1,0) {};
                \node[gauge,label=below:{$k$}] (2) at (2,0) {};
                \draw (m1)--(0)--(1);
                \draw[transform canvas={yshift=2pt}] (1)--(2);
                \draw (1)--(2);
                \draw[transform canvas={yshift=-2pt}] (1)--(2);
                \draw (1.5-0.1,0.2)--(1.5+0.1,0)--(1.5-0.1,-0.2);
            \end{tikzpicture}
        $};
    \end{tikzpicture}
    \caption{CFHM quivers $\mathsf{Q}_{k,G}$ whose $3d$ $\mathcal{N}=4$ Coulomb branch is $\mathcal{M}_{k,G}$ \cite{Cremonesi:2014xha}. We have: (a) $G=\mathrm{SU}(N)$, (b) $G=\mathrm{SO}(2N+1)$, (c) $G=\mathrm{Sp}(N)$, (d) $G=\mathrm{SO}(2N)$, (e6) $G=\mathrm{E}_6$, (e7) $G=\mathrm{E}_7$, (e8) $G=\mathrm{E}_8$, (f4) $G=\mathrm{F}_4$, (g2) $G=\mathrm{G}_2$. (a)-(d) are 3d mirror duals to Figure \ref{fig:ADHM} (a)-(d).}
    \label{fig:CFHMquivers}
\end{figure}

\paragraph{Example:} for $G=\mathrm{SU}(3)$ and $k=4$ we have the 3d mirror pair
\begin{equation}
    \vcenter{\hbox{\scalebox{1}{
    \begin{tikzpicture}
        \node[flavour,label=below:{$\mathrm{SU}(3)$}] (f) at (0,0) {};
        \node[gaugeBig, label=below:{$\mathrm{U}(4)$}] (g) at (1,0) {};
        \draw (f)--(g);
        \draw (g) to [out=315,in=45,looseness=10] (g);
        \node at (1.9,0) {Adj};
    \end{tikzpicture}}}}
    \quad\overset{\text{3d MS}}{\longleftrightarrow}\quad
    \vcenter{\hbox{\scalebox{1}{
    \begin{tikzpicture}
        \node[gauge,label=below:{$1$}] (1) at (-1,0) {};
        \node[gauge,label=below:{$4$}] (2) at (0,0) {};
        \node[gauge,label=right:{$4$}] (3) at (1,0.5) {};
        \node[gauge,label=right:{$4$}] (4) at (1,-0.5) {};
        \draw (1)--(2)--(3)--(4)--(2);
    \end{tikzpicture}}}}
\end{equation}
From now on we always take mass and Fayet-Iliopoulos parameters to be zero. As shown in \cite{deBoer:1996mp,deBoer:1996ck} we have
\begin{equation}
    \begin{tabular}{ll}
    $\mathcal{C}\left(\vcenter{\hbox{\scalebox{1}{
    \begin{tikzpicture}
        \node[gauge,label=below:{$1$}] (1) at (-1,0) {};
        \node[gauge,label=below:{$4$}] (2) at (0,0) {};
        \node[gauge,label=right:{$4$}] (3) at (1,0.5) {};
        \node[gauge,label=right:{$4$}] (4) at (1,-0.5) {};
        \draw (1)--(2)--(3)--(4)--(2);
    \end{tikzpicture}}}}
    \right)=\mathcal{M}_{4,\mathrm{SU}(3)}$ & $\mathcal{H}\left(\vcenter{\hbox{\scalebox{1}{
    \begin{tikzpicture}
        \node[gauge,label=below:{$1$}] (1) at (-1,0) {};
        \node[gauge,label=below:{$4$}] (2) at (0,0) {};
        \node[gauge,label=right:{$4$}] (3) at (1,0.5) {};
        \node[gauge,label=right:{$4$}] (4) at (1,-0.5) {};
        \draw (1)--(2)--(3)--(4)--(2);
    \end{tikzpicture}}}}
    \right)=\textnormal{Sym}^4\left(\mathbb{C}^2/\mathbb{Z}_{3}\right)$\\
    $\mathcal{C}\left(\vcenter{\hbox{\scalebox{1}{
    \begin{tikzpicture}
        \node[flavour,label=below:{$\mathrm{SU}(3)$}] (f) at (0,0) {};
        \node[gaugeBig, label=below:{$\mathrm{U}(4)$}] (g) at (1,0) {};
        \draw (f)--(g);
        \draw (g) to [out=315,in=45,looseness=10] (g);
        \node at (1.9,0) {Adj};
    \end{tikzpicture}}}}
    \right)=\textnormal{Sym}^4\left(\mathbb{C}^2/\mathbb{Z}_{3}\right)$ & $\mathcal{H}\left(\vcenter{\hbox{\scalebox{1}{
    \begin{tikzpicture}
        \node[flavour,label=below:{$\mathrm{SU}(3)$}] (f) at (0,0) {};
        \node[gaugeBig, label=below:{$\mathrm{U}(4)$}] (g) at (1,0) {};
        \draw (f)--(g);
        \draw (g) to [out=315,in=45,looseness=10] (g);
        \node at (1.9,0) {Adj};
    \end{tikzpicture}}}}
    \right)=\mathcal{M}_{4,\mathrm{SU}(3)}$
    \end{tabular}
\end{equation}
Since the rest of the paper is dedicated to study the moduli space of instantons, i.e.\ the Higgs branch of the bottom ADHM quiver, let us at this point say at least some words about the $3d$ Coulomb branch of this quiver. The $3d$ $\mathcal{N}=4$ theory can be realised on $4$ D2 branes probing a stack of $3$ coincident D6 branes in Type IIA String Theory. Lifting this system to M-theory we obtain $4$ M2 branes probing a $\mathbb{C}^2/\mathbb{Z}_3$ singularity. The moduli space of the M2 branes moving transverse to the singularity is $\textnormal{Sym}^4\left(\mathbb{C}^2/\mathbb{Z}_{3}\right)$, which is the Coulomb branch of our theory \cite{Porrati:1996xi}.
In the following we work explicitly with this example most of the time, as one can expect that our findings can easily be generalised to any $G$ and $k$.

\paragraph{Symplectic duality}

For $G=ADE$ $\mathcal{M}_{k,G}$ and $\mathrm{Sym}^k(\mathbb{C}^2/\Gamma_G)$ are proposed to be symplectic duals \cite[Remark 10.13]{Webster1407}, which is natural to expect as they are the Coulomb and Higgs branch of the same quiver, constructed in a brane system. Although we are not aware of a general theorem, in all the cases we know the number of leaves in a symplectic singularity and its symplectic dual are the same.

\section{\texorpdfstring{Towards the Hasse diagram of $\mathcal{M}_{k,G}$}{Towards the Hasse diagram}}
\label{sec:main}

In this section we develop the Hasse diagrams of the moduli space of instantons step by step. We explicitly work with the example of $\mathrm{SU}(N)$ instantons, the results can be straightforwardly applied to any other algebra.

\paragraph{Leaf counting.} In \cite[9.4 (vii)]{Webster1407} it is pointed out that the symplectic leaves of $\textnormal{Sym}^k\left(\mathbb{C}^2/\mathbb{Z}_N\right)$, $N>1$ are in bijection with partitions of $k^\prime\in\mathbb{N}$ with $k^\prime\leq k$. From symplectic duality we infer, that the number of leaves of $\mathcal{M}_{k,\mathrm{SU}(N)}$ is the same.

\subsection{Partial Higgsing}

Let us focus on the description of the instanton moduli space as the Higgs branch of an ADHM quiver, in particular Figure \ref{fig:ADHM_A}, i.e.\ a $\mathrm{U}(k)$ gauge theory with one adjoint and $N$ fundamental hypermultiplets. 

When $N \geq 2$ and $k=1$ there is a Higgsing by giving a vev to scalars in the fundamental hyper:
\begin{equation}
    \begin{tikzpicture}
        \node (1) at (0,0) {$\begin{tikzpicture}
                \node[flavour,label=left:{$\mathrm{SU}(N)$}] (f) at (0,1) {};
                \node[gaugeBig, label=left:{$\mathrm{U}(1)$}] (g) at (0,0) {};
                \draw (f)--(g);
                \draw (g) to [out=225,in=315,looseness=10] (g);
                \node at (0,-1) {Adj};
        \end{tikzpicture}$};
        \node (2) at (4,0) {free hypers};
        \draw[->] (1)--(2);
    \end{tikzpicture}\;.
    \label{eq:fundHiggs}
\end{equation}
Note that for $N=1$ the arrow (\ref{eq:fundHiggs}) becomes an identity, and there is no transition. 

Furthermore, for $N \geq 1$ and $k \geq 2$ we can use representation theory to describe the partial Higgs mechanism, using the branching rules
\begin{equation}
    \begin{split}
        \mathrm{U}(k)&\rightarrow \mathrm{U}(k-l)\times \mathrm{U}(l)\\
        \textnormal{adj}_k&\mapsto \textnormal{adj}_{k-l}+\textnormal{adj}_{l}+\textnormal{fund}_{k-l}\overline{\textnormal{fund}}_{l}+\overline{\textnormal{fund}}_{k-l}\textnormal{fund}_{l}\\
        \textnormal{fund}_k&\mapsto \textnormal{fund}_{k-l}+\textnormal{fund}_{l}\,,
    \end{split}
\end{equation}
where $\textnormal{adj}_k$ and $\textnormal{fund}_k$ denote the adjoint and fundamental representation of $\mathrm{U}(k)$.
There seems to be a partial Higgsing by giving a vev to scalars in the adjoint hyper:
\begin{equation}
    \begin{tikzpicture}
        \node (1) at (0,0) {$\begin{tikzpicture}
                \node[flavour,label=left:{$A_{N-1}$}] (f) at (0,1) {};
                \node[gaugeBig, label=left:{$\mathrm{U}(k)$}] (g) at (0,0) {};
                \draw (f)--(g);
                \draw (g) to [out=225,in=315,looseness=10] (g);
                \node at (0,-1) {Adj};
        \end{tikzpicture}$};
        \node (2) at (4,0) {$\begin{tikzpicture}
                \node[flavour,label=left:{$A_{N-1}$}] (f) at (0,1) {};
                \node[gaugeBig, label=left:{$\mathrm{U}(k-l)$}] (g) at (0,0) {};
                \draw (f)--(g);
                \draw (g) to [out=225,in=315,looseness=10] (g);
                \node at (0,-1) {Adj};
        \end{tikzpicture}$ $\begin{tikzpicture}
                \node[flavour,label=right:{$A_{N-1}$}] (f) at (0,1) {};
                \node[gaugeBig, label=right:{$\mathrm{U}(l)$}] (g) at (0,0) {};
                \draw (f)--(g);
                \draw (g) to [out=225,in=315,looseness=10] (g);
                \node at (0,-1) {Adj};
        \end{tikzpicture}$};
        \draw[->] (1)--(2);
    \end{tikzpicture}\,.
    \label{eq:adjHiggs}
\end{equation}
Let us consider the case $k=2$. We have the Higgsing $\mathrm{U}(2) \rightarrow \mathrm{U}(1)\times \mathrm{U}(1)$, but there is a caveat. The Weyl group of $\mathrm{U}(2)$ is $\mathbb{Z}_2$. The action of this $\mathbb{Z}_2$ swaps the two $\mathrm{U}(1)$s inside the $\mathrm{U}(2)$. Since this $\mathbb{Z}_2$ is a gauge symmetry before Higgsing, and it is not broken by the adjoint Higgsing \eqref{eq:adjHiggs}, it must remain as a gauge symmetry after Higgsing. Hence it would be more precise to write something like:
\begin{equation}
    \begin{tikzpicture}
        \node (1) at (0,0) {$\begin{tikzpicture}
                \node[flavour,label=left:{$A_{N-1}$}] (f) at (0,1) {};
                \node[gaugeBig, label=left:{$\mathrm{U}(2)$}] (g) at (0,0) {};
                \draw (f)--(g);
                \draw (g) to [out=225,in=315,looseness=10] (g);
                \node at (0,-1) {Adj};
        \end{tikzpicture}$};
        \node (2) at (4,0) {$\begin{tikzpicture}
                \node[flavour,label=left:{$A_{N-1}$}] (f) at (0,1) {};
                \node[gaugeBig, label=left:{$\mathrm{U}(1)$}] (g) at (0,0) {};
                \draw (f)--(g);
                \draw (g) to [out=225,in=315,looseness=10] (g);
                \node at (0,-1) {Adj};
                \node[flavour,label=right:{$A_{N-1}$}] (f1) at (1.5,1) {};
                \node[gaugeBig, label=right:{$\mathrm{U}(1)$}] (g1) at (1.5,0) {};
                \draw (f1)--(g1);
                \draw (g1) to [out=225,in=315,looseness=10] (g1);
                \node at (1.5,-1) {Adj};
                \draw[<->] (0.3,0)--(1.2,0);
                \node at (0.75,-0.3) {$\mathbb{Z}_2$};
        \end{tikzpicture}$};
        \draw[->] (1)--(2);
    \end{tikzpicture}\,.
\end{equation}
The $\mathbb{Z}_2$ has a drastic effect on the Higgs branch, and its Hasse diagram. Based on the Higgsing \eqref{eq:fundHiggs} we have:
\begin{equation}
    \mathcal{H}\left(\vcenter{\hbox{\scalebox{1}{
    \begin{tikzpicture}
                \node[flavour,label=left:{$A_{N-1}$}] (f) at (0,1) {};
                \node[gaugeBig, label=left:{$\mathrm{U}(1)$}] (g) at (0,0) {};
                \draw (f)--(g);
                \draw (g) to [out=225,in=315,looseness=10] (g);
                \node at (0,-1) {Adj};
                \node[flavour,label=right:{$A_{N-1}$}] (f1) at (1.5,1) {};
                \node[gaugeBig, label=right:{$\mathrm{U}(1)$}] (g1) at (1.5,0) {};
                \draw (f1)--(g1);
                \draw (g1) to [out=225,in=315,looseness=10] (g1);
                \node at (1.5,-1) {Adj};
        \end{tikzpicture}}}}\right)\textnormal{ has Hasse diagram }\vcenter{\hbox{\scalebox{1}{
    \begin{tikzpicture}
                \node[hasse] (1) at (0,0) {};
                \node[hasse] (2) at (-1,1) {};
                \node[hasse] (3) at (1,1) {};
                \node[hasse] (4) at (0,2) {};
                \draw (1)--(2)--(4)--(3)--(1);
                \node at (-1,0.5) {$a_{N-1}$};
                \node at (-1,1.5) {$a_{N-1}$};
                \node at (1,0.5) {$a_{N-1}$};
                \node at (1,1.5) {$a_{N-1}$};
        \end{tikzpicture}}}}\,.
\end{equation}
However, if there is a $\mathbb{Z}_2$ gauge symmetry then we cannot distinguish which $\mathrm{U}(1)$ is partially Higgsed, and therefore there is only one middle leaf:
\begin{equation}
    \mathcal{H}\left(\vcenter{\hbox{\scalebox{1}{
    \begin{tikzpicture}
                \node[flavour,label=left:{$A_{N-1}$}] (f) at (0,1) {};
                \node[gaugeBig, label=left:{$\mathrm{U}(1)$}] (g) at (0,0) {};
                \draw (f)--(g);
                \draw (g) to [out=225,in=315,looseness=10] (g);
                \node at (0,-1) {Adj};
                \node[flavour,label=right:{$A_{N-1}$}] (f1) at (1.5,1) {};
                \node[gaugeBig, label=right:{$\mathrm{U}(1)$}] (g1) at (1.5,0) {};
                \draw (f1)--(g1);
                \draw (g1) to [out=225,in=315,looseness=10] (g1);
                \node at (1.5,-1) {Adj};
                \draw[<->] (0.3,0)--(1.2,0);
                \node at (0.75,-0.3) {$\mathbb{Z}_2$};
        \end{tikzpicture}}}}\right)\textnormal{ has Hasse diagram }\vcenter{\hbox{\scalebox{1}{
    \begin{tikzpicture}
                \node[hasse] (1) at (0,0) {};
                \node[hasse] (2) at (0,1) {};
                \node[hasse] (4) at (0,2) {};
                \draw (1)--(2)--(4);
                \node at (-0.4,0.5) {$a_{N-1}$};
                \node at (-0.4,1.5) {$a_{N-1}$};
        \end{tikzpicture}}}}\,.
\end{equation}
This behaviour can be generalised to higher ranks.

\begin{figure}
    \centering
     \begin{subfigure}[b]{0.19\textwidth}
         \centering
             \begin{tikzpicture}
                \node (4) at (0,0) {$4$};
                \node (31) at (0.5,1) {$31$};
                \node (22) at (-0.5,1) {$22$};
                \node (211) at (0,2) {$211$};
                \node (1111) at (0,3) {$1111$};
                \draw (4)--(31)--(211)--(1111) (4)--(22)--(211);
            \end{tikzpicture}
         \caption{$k=4$}
         \label{fig:Higgsing04}
     \end{subfigure}
     \begin{subfigure}[b]{0.19\textwidth}
         \centering
             \begin{tikzpicture}
                \node (3) at (0,1.5) {$3$};
                \node (21) at (0,2.5) {$21$};
                \node (111) at (0,3.5) {$111$};
                \draw (3)--(21)--(111);
                \node at (0,0) {};
            \end{tikzpicture}
         \caption{$k=3$}
         \label{Higgsing03}
     \end{subfigure}
     \begin{subfigure}[b]{0.19\textwidth}
         \centering
             \begin{tikzpicture}
                \node (2) at (0,3) {$2$};
                \node (11) at (0,4) {$11$};
                \draw (2)--(11);
                \node at (0,0) {};
            \end{tikzpicture}
         \caption{$k=2$}
         \label{Higgsing02}
     \end{subfigure}
     \begin{subfigure}[b]{0.19\textwidth}
         \centering
             \begin{tikzpicture}
                \node (1) at (0,4.5) {$1$};
                \node at (0,0) {};
            \end{tikzpicture}
         \caption{$k=1$}
         \label{Higgsing01}
     \end{subfigure}
     \begin{subfigure}[b]{0.19\textwidth}
         \centering
             \begin{tikzpicture}
\node at (0,0) {};
            \end{tikzpicture}
         \caption{$k=0$}
         \label{Higgsing00}
     \end{subfigure}
    \caption{The partial Higgsing pattern for $\mathrm{U}(k)$ with one adjoint and $N=1$ fundamental hypermultiplet, for $4\geq k\geq 0$. The partitions of $k$ denote the residual gauge group.     }
    \label{fig:Higgsing0}
\end{figure}
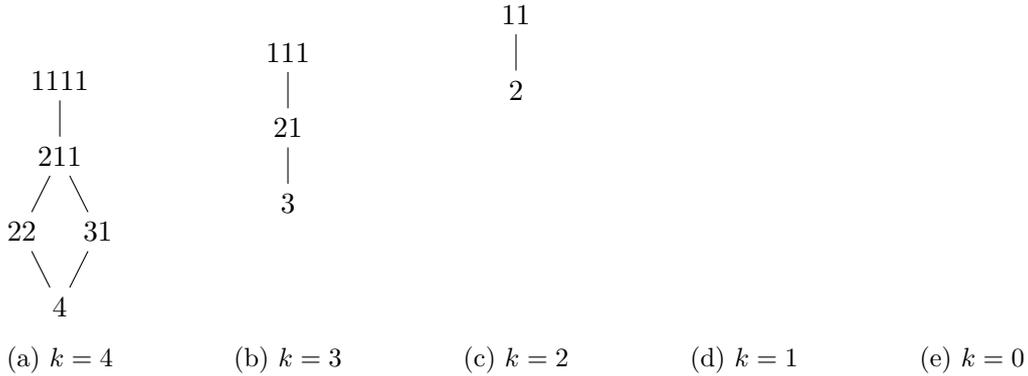

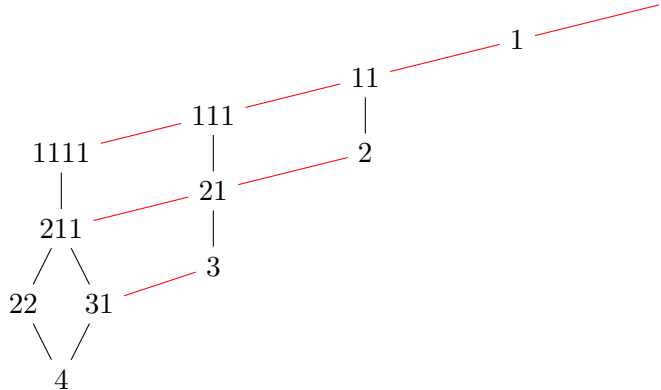
\begin{figure}
    \centering
             \begin{tikzpicture}
                \node (4) at (0,0) {$4$};
                \node (31) at (0.5,1) {$31$};
                \node (22) at (-0.5,1) {$22$};
                \node (211) at (0,2) {$211$};
                \node (1111) at (0,3) {$1111$};
                \draw (4)--(31)--(211)--(1111) (4)--(22)--(211);
                \node (3) at (2,1.5) {$3$};
                \node (21) at (2,2.5) {$21$};
                \node (111) at (2,3.5) {$111$};
                \draw (3)--(21)--(111);
                \node (2) at (4,3) {$2$};
                \node (11) at (4,4) {$11$};
                \draw (2)--(11);
                \node (1) at (6,4.5) {$1$};
                \node (0) at (8,5) {};
                \draw[red] (31)--(3) (211)--(21)--(2) (1111)--(111)--(11)--(1)--(0);
            \end{tikzpicture}
    \caption{The partial Higgsing pattern for $\mathrm{U}(4)$ with an adjoint and $N>1$ fundamental hypermultiplets. The partitions of $0\leq k'\leq k=4$ denote the residual gauge group.
    The black transitions correspond to adjoint Higgsings \eqref{eq:adjHiggs} leaving the rank intact, the red transitions correspond to fundamental Higgsings \eqref{eq:fundHiggs} of $\mathrm{U}(1)$.}
    \label{fig:Higgsing4}
\end{figure}

Since the fundamental Higgs mechanism (\ref{eq:fundHiggs}) is only available when $N \geq 2$, the shape of the Hasse diagrams depends on the value of $N$: 
\begin{itemize}
    \item For $N=1$, we have 
    \begin{equation}
    \mathcal{H}\left(\vcenter{\hbox{\scalebox{1}{
    \begin{tikzpicture}
        \node[flavour,label=below:{$1$}] (f) at (0,0) {};
        \node[gaugeBig, label=below:{$\mathrm{U}(k)$}] (g) at (1,0) {};
        \draw (f)--(g);
        \draw (g) to [out=315,in=45,looseness=10] (g);
        \node at (1.9,0) {Adj};
    \end{tikzpicture}}}}\right)=\textnormal{Sym}^k\left(\mathbb{C}^2\right)\;.
\end{equation}
We can obtain the Hasse diagram, displayed in Figure \ref{fig:Higgsing0} for various $k$, from adjoint Higgsings \eqref{eq:adjHiggs}.
\item For $N \geq 2$ we have to take into account both Higgsings \eqref{eq:adjHiggs} and \eqref{eq:fundHiggs}. For $k=4$ the Hasse diagram is given in Figure \ref{fig:Higgsing4}. Working in terms of partitions of $k^\prime\leq k$ makes it relatively easy to generate the Hasse diagram for any $k$. For large $k$ the Hasse diagram becomes quite involved.
\end{itemize}

From the Higgsing pattern of our ADHM quiver we have so far obtained the partial order of inclusions of closures of the symplectic leaves in the moduli space of instantons. The next step is to identify the elementary slices. For the Higgsings \eqref{eq:fundHiggs}, i.e.\ the red lines in Figure \ref{fig:Higgsing4}, the associated elementary slice is $a_{N-1}$, as this is a small instanton transition. We can generalise this to any other algebra. The nature of the elementary slices associated to the Higgsing \eqref{eq:adjHiggs}, i.e.\ the black lines, is addressed below.

\subsection{Branes within Branes}

The string theory description of instantons as branes within branes mentioned in Section \ref{sec:generalities} allows us to get some understanding of the moduli space Hasse diagram. Let us consider the specific example of $k$ D3 branes probing a coincident stack of $N>1$ D7 branes. The moduli space of instantons is realised when all $k$ D3 branes are inside the D7 brane stack. On a general point of this moduli space (i.e. on the highest leaf) all the D3 branes have a non-zero size. In order to move to a lower leaf in the moduli space we can shrink a D3 brane to zero size, corresponding to the fundamental Higgsing \eqref{eq:fundHiggs}, an $a_{N-1}$ transition. This requires the tuning of $N-1$ moduli. Since all $k$ D3 branes are identical there is only one lower leaf reached by such a transition. We can now continue and shrink a second D3, again corresponding to an $a_{N-1}$ transition, reaching an even lower leaf. Now we have two possibilities:
\begin{enumerate}
    \item We can shrink a third D3 brane.
    \item We can take the two zero size D3 branes and move them on top of each other. This corresponds to the adjoint Higgsing \eqref{eq:adjHiggs}.
\end{enumerate}
The two transitions bring us to two different lower leaves. Let us focus on option 2. Since the two D3 branes we collide are identical this transition corresponds to $\mathbb{C}^2/\mathbb{Z}_2$.

In general, for $k$ D3 branes, we can tune all $k$ D3 branes to zero size, leading to $k$ $a_{N-1}$ transitions. Whenever we have at least two zero size D3 branes, we can start moving them on top of each other. When we move a third D3 brane on top of two coincident D3 branes, or more generally two D3 brane stacks containing a different number of D3 branes, it is not immediately clear what the corresponding transition is. However, whenever we move two stacks containing the same number of D3 branes together, we can argue the corresponding slice to be $\mathbb{C}^2/\mathbb{Z}_2$. In Figure \ref{fig:shrinkingBranes} we produce the Hasse diagram for $4$ D3 branes inside a stack of $N$ D7 branes.

\begin{figure}
    \centering
             \begin{tikzpicture}
                \node (4) at (-0.3,0) {$\begin{tikzpicture}
                    \node[shrinky] (4x1) at (0,0) {};
                    \node[shrinky] (4x2) at (0.3,0) {};
                    \node[shrinky] (4x3) at (0.6,0) {};
                    \node[shrinky] (4x4) at (0.9,0) {};
                    \draw \convexpath{4x1,4x4}{3pt};
                \end{tikzpicture}
                $};
                \node (31) at (0.7-0.3,1) {$\begin{tikzpicture}
                    \node[shrinky] (31x1) at (0,0) {};
                    \node[shrinky] (31x2) at (0.3,0) {};
                    \node[shrinky] (31x3) at (0.6,0) {};
                    \node[shrinky] (31x4) at (0.9,0) {};
                    \draw \convexpath{31x1,31x3}{3pt};
                \end{tikzpicture}
                $};
                \node (22) at (-0.7-0.3,1) {$\begin{tikzpicture}
                    \node[shrinky] (22x1) at (0,0) {};
                    \node[shrinky] (22x2) at (0.3,0) {};
                    \node[shrinky] (22x3) at (0.6,0) {};
                    \node[shrinky] (22x4) at (0.9,0) {};
                    \draw \convexpath{22x1,22x2}{3pt};
                    \draw \convexpath{22x3,22x4}{3pt};
                \end{tikzpicture}
                $};
                \node (211) at (0-0.3,2) {$\begin{tikzpicture}
                    \node[shrinky] (211x1) at (0,0) {};
                    \node[shrinky] (211x2) at (0.3,0) {};
                    \node[shrinky] (211x3) at (0.6,0) {};
                    \node[shrinky] (211x4) at (0.9,0) {};
                    \draw \convexpath{211x1,211x2}{3pt};
                \end{tikzpicture}
                $};
                \node (1111) at (0-0.3,3) {$\begin{tikzpicture}
                    \node[shrinky] (1111x1) at (0,0) {};
                    \node[shrinky] (1111x2) at (0.3,0) {};
                    \node[shrinky] (1111x3) at (0.6,0) {};
                    \node[shrinky] (1111x4) at (0.9,0) {};
                \end{tikzpicture}
                $};
                \draw (4)--(31)--(211)--(1111) (4)--(22)--(211);
                \node (3) at (2,1.5) {$\begin{tikzpicture}
                    \node[shrinky] (3x1) at (0,0) {};
                    \node[shrinky] (3x2) at (0.3,0) {};
                    \node[shrinky] (3x3) at (0.6,0) {};
                    \node[sized] (3x4) at (0.9,0) {};
                    \draw \convexpath{3x1,3x3}{3pt};
                \end{tikzpicture}
                $};
                \node (21) at (2,2.5) {$\begin{tikzpicture}
                    \node[shrinky] (21x1) at (0,0) {};
                    \node[shrinky] (21x2) at (0.3,0) {};
                    \node[shrinky] (21x3) at (0.6,0) {};
                    \node[sized] (21x4) at (0.9,0) {};
                    \draw \convexpath{21x1,21x2}{3pt};
                \end{tikzpicture}
                $};
                \node (111) at (2,3.5) {$\begin{tikzpicture}
                    \node[shrinky] (111x1) at (0,0) {};
                    \node[shrinky] (111x2) at (0.3,0) {};
                    \node[shrinky] (111x3) at (0.6,0) {};
                    \node[sized] (111x4) at (0.9,0) {};
                \end{tikzpicture}
                $};
                \draw (3)--(21)--(111);
                \node (2) at (4,3) {$\begin{tikzpicture}
                    \node[shrinky] (2x1) at (0,0) {};
                    \node[shrinky] (2x2) at (0.3,0) {};
                    \node[sized] (2x3) at (0.6,0) {};
                    \node[sized] (2x4) at (0.9,0) {};
                    \draw \convexpath{2x1,2x2}{3pt};
                \end{tikzpicture}
                $};
                \node (11) at (4,4) {$\begin{tikzpicture}
                    \node[shrinky] (11x1) at (0,0) {};
                    \node[shrinky] (11x2) at (0.3,0) {};
                    \node[sized] (11x3) at (0.6,0) {};
                    \node[sized] (11x4) at (0.9,0) {};
                \end{tikzpicture}
                $};
                \draw (2)--(11);
                \node (1) at (6,4.5) {$\begin{tikzpicture}
                    \node[shrinky] (1x1) at (0,0) {};
                    \node[sized] (1x2) at (0.3,0) {};
                    \node[sized] (1x3) at (0.6,0) {};
                    \node[sized] (1x4) at (0.9,0) {};
                \end{tikzpicture}
                $};
                \node (0) at (8,5) {$\begin{tikzpicture}
                    \node[sized] (0x1) at (0,0) {};
                    \node[sized] (0x2) at (0.3,0) {};
                    \node[sized] (0x3) at (0.6,0) {};
                    \node[sized] (0x4) at (0.9,0) {};
                \end{tikzpicture}
                $};
                \draw[red] (31)--(3) (211)--(21)--(2) (1111)--(111)--(11)--(1)--(0);
            \end{tikzpicture}
    \caption[]{The Hasse diagram obtained from shrinking (red) and colliding (black) $4$ D3 branes inside a stack of $N$ D7 branes. We draw D3 branes with a size as circles $\vcenter{\hbox{\scalebox{1}{\begin{tikzpicture}
        \node[sized] at (0,0) {};
    \end{tikzpicture}}}}$, D3 branes of zero size as dots $\vcenter{\hbox{\scalebox{1}{\begin{tikzpicture}
        \node[shrinky] at (0,0) {};
    \end{tikzpicture}}}}$, and denote coincident stacks of D3 branes by drawing a line around them $\vcenter{\hbox{\scalebox{1}{\begin{tikzpicture}
        \node[shrinky] (1) at (0,0) {};
        \node[shrinky] (2) at (0.3,0) {};
        \draw \convexpath{1,2}{3pt};
    \end{tikzpicture}}}}$.}
    \label{fig:shrinkingBranes}
\end{figure}

Clearly the Hasse diagram produced from partial Higgsings matches the Hasse diagram using brane arguments. This is no surprise.

\subsection{Quiver Subtraction}

In this section we use quiver subtraction on CFHM quivers (Figure \ref{fig:CFHMquivers}) to obtain the Hasse diagrams of the previous sections. In order to do this, we have to generalize the quiver subtraction algorithm of \cite{Cabrera:2018ann,Bourget:2019aer}. For this it is useful to revisit the way we think about quiver subtraction. As a first example, we consider the CFHM quiver for the moduli space of $2$ $\mathrm{SU}(3)$ instantons.

\paragraph{Contracting Loops.} As pointed out in \cite[A.2]{Bourget:2019aer} subtracting an affine diagram from a quiver can be viewed as contracting a loop\footnote{Here we refer to any affine Dynkin quiver as a loop.}:
\begin{equation}
    \begin{tikzpicture}
        \node (a) at (0,0) {
    $\begin{tikzpicture}
        \node[gauge,label=below:{$1$}] (1) at (-1,0) {};
        \node[gauge,label=below:{$2$}] (2) at (0,0) {};
        \node[gauge,label=right:{$2$}] (3) at (1,0.5) {};
        \node[gauge,label=right:{$2$}] (4) at (1,-0.5) {};
        \draw (1)--(2)--(3)--(4)--(2);
    \end{tikzpicture}$};
        \node (b) at (3,-3) {
    $\begin{tikzpicture}
        \node[gauge,label=below:{$1$}] (1) at (-1,0) {};
        \node[gauge,label=below:{$1$}] (2) at (0,0) {};
        \node[gauge,label=right:{$1$}] (3) at (1,0.5) {};
        \node[gauge,label=right:{$1$}] (4) at (1,-0.5) {};
        \draw (1)--(2)--(3)--(4)--(2);
        \begin{scope}[rotate around={60:(-1,0)}]
        \node[gauge,label=below:{$1$}] (5) at (0,0) {};
        \node[gauge,label=right:{$1$}] (6) at (1,0.5) {};
        \node[gauge,label=right:{$1$}] (7) at (1,-0.5) {};
        \draw (5)--(6)--(7)--(5);
        \end{scope}
        \draw (5)--(1);
        \draw \convexpath{5,6,7}{0.6cm};
        \draw[dashed] \convexpath{2,3,4}{0.6cm};
    \end{tikzpicture}$};
        \node (c) at (0,-6) {
    $\begin{tikzpicture}
        \node[gauge,label=below:{$1$}] (1) at (-1,0) {};
        \node[gauge,label=below:{$1$}] (2) at (0,0) {};
        \node[gauge,label=right:{$1$}] (3) at (1,0.5) {};
        \node[gauge,label=right:{$1$}] (4) at (1,-0.5) {};
        \draw (1)--(2)--(3)--(4)--(2);
        \begin{scope}[rotate around={60:(-1,0)}]
        \node[gauge,label=below:{$1$}] (5) at (0,0) {};
        \end{scope}
        \draw (5)--(1);
        \draw (5) circle (0.6cm);
        \draw[dashed] \convexpath{2,3,4}{0.6cm};
    \end{tikzpicture}$};
    \draw[->] (a) .. controls (1,-3) .. (c);
    \node at (0,-3) {$-a_2$};
    \end{tikzpicture}
    \label{eq:ContractingLoop}
\end{equation}
In \eqref{eq:ContractingLoop} however, the `loop' is present multiple times in the original quiver, visible both inside the solid outline and the dashed outline. It was pointed out in \cite[C]{Bourget:2020mez}, that in such a case the quiver has to be {\color{purple}\emph{decorated}}. The decoration encodes the fact, that any of the two loops in \eqref{eq:ContractingLoop} could have been contracted, as they are identical. After contracting one loop in \eqref{eq:ContractingLoop} one can contract the second loop:
\begin{equation}
    \begin{tikzpicture}
        \node (a) at (0,0) {
    $\begin{tikzpicture}
        \node[gauge,label=below:{$1$}] (1) at (-1,0) {};
        \node[gauge,label=below:{$2$}] (2) at (0,0) {};
        \node[gauge,label=right:{$2$}] (3) at (1,0.5) {};
        \node[gauge,label=right:{$2$}] (4) at (1,-0.5) {};
        \draw (1)--(2)--(3)--(4)--(2);
    \end{tikzpicture}$};
        \node (b) at (3,-3) {
    $\begin{tikzpicture}
        \node[gauge,label=below:{$1$}] (1) at (-1,0) {};
        \node[gauge,label=below:{$1$}] (2) at (0,0) {};
        \node[gauge,label=right:{$1$}] (3) at (1,0.5) {};
        \node[gauge,label=right:{$1$}] (4) at (1,-0.5) {};
        \draw (1)--(2)--(3)--(4)--(2);
        \begin{scope}[rotate around={60:(-1,0)}]
        \node[gauge,label=below:{$1$}] (5) at (0,0) {};
        \node[gauge,label=right:{$1$}] (6) at (1,0.5) {};
        \node[gauge,label=right:{$1$}] (7) at (1,-0.5) {};
        \draw (5)--(6)--(7)--(5);
        \end{scope}
        \draw (5)--(1);
        \draw[purple] \convexpath{5,6,7}{0.6cm};
        \draw[purple] \convexpath{2,3,4}{0.6cm};
    \end{tikzpicture}$};
        \node (c) at (0,-6) {
    $\begin{tikzpicture}
        \node[gauge,label=below:{$1$}] (1) at (-1,0) {};
        \node[gauge,label=below:{$1$}] (2) at (0,0) {};
        \node[gauge,label=right:{$1$}] (3) at (1,0.5) {};
        \node[gauge,label=right:{$1$}] (4) at (1,-0.5) {};
        \draw (1)--(2)--(3)--(4)--(2);
        \begin{scope}[rotate around={60:(-1,0)}]
        \node[gauge,label=below:{$1$}] (5) at (0,0) {};
        \end{scope}
        \draw (5)--(1);
        \draw[purple] (5) circle (0.6cm);
        \draw[purple] \convexpath{2,3,4}{0.6cm};
    \end{tikzpicture}$};
    \draw[->] (a) .. controls (1,-3) .. (c);
    \node at (0,-3) {$-a_2$};
        \node (d) at (3,-9) {
    $\begin{tikzpicture}
        \node[gauge,label=below:{$1$}] (1) at (-1,0) {};
        \node[gauge,label=below:{$1$}] (2) at (0,0) {};
        \node[gauge,label=right:{$1$}] (3) at (1,0.5) {};
        \node[gauge,label=right:{$1$}] (4) at (1,-0.5) {};
        \draw (1)--(2)--(3)--(4)--(2);
        \begin{scope}[rotate around={60:(-1,0)}]
        \node[gauge,label=below:{$1$}] (5) at (0,0) {};
        \end{scope}
        \draw (5)--(1);
        \draw[purple] (5) circle (0.6cm);
        \draw[purple] \convexpath{2,3,4}{0.6cm};
    \end{tikzpicture}$};
        \node (e) at (0,-12) {
    $\begin{tikzpicture}
        \node[gauge,label=below:{$1$}] (1) at (-1,0) {};
        \node[gauge,label=below:{$1$}] (2) at (0,0) {};
        \draw (1)--(2);
        \begin{scope}[rotate around={60:(-1,0)}]
        \node[gauge,label=below:{$1$}] (5) at (0,0) {};
        \end{scope}
        \draw (5)--(1);
        \draw[purple] (5) circle (0.6cm);
        \draw[purple] (2) circle (0.6cm);
    \end{tikzpicture}$};
    \draw[->] (c) .. controls (1,-9) .. (e);
    \node at (0,-9) {$-a_2$};
    \end{tikzpicture}
    \label{eq:ContractingLoop2}
\end{equation}
The final quiver in \eqref{eq:ContractingLoop2} contains two $\mathrm{U}(1)$ nodes decorated with the same colour. We have contracted two identical loops. Hence there is a discrete symmetry exchanging the two $\mathrm{U}(1)$s, which is gauged. The final quiver in \eqref{eq:ContractingLoop2} is conjectured to be equivalent to:
\begin{equation}
    \vcenter{\hbox{\scalebox{1}{\begin{tikzpicture}
        \node[gauge,label=below:{$1$}] (1) at (-1,0) {};
        \node[gauge,label=below:{$1$}] (2) at (0,0) {};
        \draw (1)--(2);
        \begin{scope}[rotate around={60:(-1,0)}]
        \node[gauge,label=below:{$1$}] (5) at (0,0) {};
        \end{scope}
        \draw (5)--(1);
        \draw[purple] (5) circle (0.6cm);
        \draw[purple] (2) circle (0.6cm);
    \end{tikzpicture}}}}=
    \vcenter{\hbox{\scalebox{1}{\begin{tikzpicture}
        \node[gauge,label=below:{$1$}] (1) at (0,0) {};
        \node[gauge,label=left:{$2$}] (2) at (0,1) {};
        \draw (1)--(2);
        \draw (2) to [out=45,in=135,looseness=10] (2);
    \end{tikzpicture}}}}\;.
    \label{eq:DecorateIsDiscreteGauge}
\end{equation}
If we were to take the left quiver in \eqref{eq:DecorateIsDiscreteGauge} without decoration, its Coulomb branch would be $\mathbb{H}^2$. Since the free factor in $\mathcal{M}_{k,G}$ is $\mathbb{H}$ the closure of a symplectic leaf cannot contain an $\mathbb{H}^2$ factor. Hence the quiver without decoration does not describe the closure of a leaf in $\mathcal{M}_{2,\mathrm{SU}(3)}$.

Gauging the $\mathbb{Z}_2$ implied by the decoration leads us to expect that the Coulomb branch of the decorated quiver is $\mathbb{H}^2/\mathbb{Z}_2$. The Coulomb branch of the right quiver in \eqref{eq:DecorateIsDiscreteGauge} is $\mathbb{H}\times A_1$, which is indeed a $\mathbb{Z}_2$ quotient of $\mathbb{H}^2$. This time we have the expected free factor $\mathbb{H}$.

Since the $A_1$ factor is singular, there should be a further `quiver subtraction' of an $A_1$ slice leading to a quiver with Coulomb branch $\mathbb{H}$. In order to do this we have to figure out how to deal with identical contracted loops.

\paragraph{Merging identical contracted Loops.} Before we dive right in, it is instructive to consider the following quiver and Coulomb branch \cite{Kraft1982,Hanany:2018dvd,Bourget:2020bxh}:
\begin{equation}
    \mathcal{C}\left(\vcenter{\hbox{\scalebox{1}{
    \begin{tikzpicture}
        \node[gauge,label=below:{$1$}] (1) at (0,0) {};
        \node[gauge,label=below:{$2$}] (2) at (1,0) {};
        \node[gauge,label=left:{$1$}] (3) at (1,1) {};
        \node[gauge,label=below:{$2$}] (4) at (2,0) {};
        \draw (1)--(2)--(3) (4)--(2);
        \draw (4) to [out=315,in=45,looseness=10] (4);
    \end{tikzpicture}}}}
    \right)=\overline{\textnormal{n.min B}{}_3}\qquad\textnormal{with Hasse diagram}\qquad
    \vcenter{\hbox{\scalebox{1}{
    \begin{tikzpicture}
        \node[hasse] (1) at (0,0) {};
        \node[hasse] (2) at (0,1) {};
        \node[hasse] (3) at (0,2) {};
        \draw (1)--(2)--(3);
        \node at (-0.4,0.5) {$b_3$};
        \node at (-0.4,1.5) {$A_1$};
    \end{tikzpicture}}}}\;.
    \label{eq:NMinB3}
\end{equation}
The quiver for the bottom slice is
\begin{equation}
    \mathcal{C}\left(\vcenter{\hbox{\scalebox{1}{
    \begin{tikzpicture}
        \node[gauge,label=below:{$1$}] (1) at (0,0) {};
        \node[gauge,label=below:{$2$}] (2) at (1,0) {};
        \node[gauge,label=left:{$1$}] (3) at (1,1) {};
        \node[gauge,label=below:{$1$}] (4) at (2,0) {};
        \draw (1)--(2)--(3);
        \draw[transform canvas={yshift=1pt}] (2)--(4);
        \draw[transform canvas={yshift=-1pt}] (2)--(4);
        \draw (1.5-0.1,0.2)--(1.5+0.1,0)--(1.5-0.1,-0.2);
    \end{tikzpicture}}}}
    \right)=\overline{\textnormal{min B}{}_3}\qquad\textnormal{with Hasse diagram}\qquad
    \vcenter{\hbox{\scalebox{1}{
    \begin{tikzpicture}
        \node[hasse] (1) at (0,0) {};
        \node[hasse] (2) at (0,1) {};
        \draw (1)--(2);
        \node at (-0.4,0.5) {$b_3$};
    \end{tikzpicture}}}}\;.
    \label{eq:MinB3}
\end{equation}
We may replace the $\mathrm{U}(2)$ node with an adjoint in \eqref{eq:NMinB3} with two $\mathrm{U}(1)$s decorated with the same colour:
\begin{equation}
    \vcenter{\hbox{\scalebox{1}{
    \begin{tikzpicture}
        \node[gauge,label=below:{$1$}] (1) at (0,0) {};
        \node[gauge,label=below:{$2$}] (2) at (1,0) {};
        \node[gauge,label=left:{$1$}] (3) at (1,1) {};
        \node[gauge,label=below:{$2$}] (4) at (2,0) {};
        \draw (1)--(2)--(3) (4)--(2);
        \draw (4) to [out=315,in=45,looseness=10] (4);
    \end{tikzpicture}}}}=
    \vcenter{\hbox{\scalebox{1}{
    \begin{tikzpicture}
        \node[gauge,label=below:{$1$}] (1) at (0,0) {};
        \node[gauge,label=below:{$2$}] (2) at (1,0) {};
        \node[gauge,label=left:{$1$}] (3) at (1,1) {};
        \node[gauge,label=right:{$1$}] (4u) at (2,0.5) {};
        \node[gauge,label=right:{$1$}] (4d) at (2,-0.5) {};
        \draw (1)--(2)--(3) (4u)--(2)--(4d);
        \draw[purple] (4u) circle (0.5cm);
        \draw[purple] (4d) circle (0.5cm);
    \end{tikzpicture}}}}
    \label{eq:NMinB3Decorated}
\end{equation}
In order to obtain the quiver in \eqref{eq:MinB3} we should merge the two decorated $\mathrm{U}(1)$s in \eqref{eq:NMinB3Decorated} into a single `short' $\mathrm{U}(1)$, corresponding to an $A_1$ transition:
\begin{equation}
    \begin{tikzpicture}
        \node (a) at (0,0) {
    $\begin{tikzpicture}
        \node[gauge,label=below:{$1$}] (1) at (0,0) {};
        \node[gauge,label=below:{$2$}] (2) at (1,0) {};
        \node[gauge,label=left:{$1$}] (3) at (1,1) {};
        \node[gauge,label=right:{$1$}] (4u) at (2,0.5) {};
        \node[gauge,label=right:{$1$}] (4d) at (2,-0.5) {};
        \draw (1)--(2)--(3) (4u)--(2)--(4d);
        \draw[purple] (4u) circle (0.5cm);
        \draw[purple] (4d) circle (0.5cm);
    \end{tikzpicture}$};
        \node (c) at (0,-4) {
    $\begin{tikzpicture}
        \node[gauge,label=below:{$1$}] (1) at (0,0) {};
        \node[gauge,label=below:{$2$}] (2) at (1,0) {};
        \node[gauge,label=left:{$1$}] (3) at (1,1) {};
        \node[gauge,label=below:{$1$}] (4) at (2,0) {};
        \draw (1)--(2)--(3);
        \draw[purple] (4) circle (0.5cm);
        \draw[transform canvas={yshift=1pt}] (2)--(4);
        \draw[transform canvas={yshift=-1pt}] (2)--(4);
        \draw (1.5-0.1,0.2)--(1.5+0.1,0)--(1.5-0.1,-0.2);
    \end{tikzpicture}$};
    \draw[->] (a) .. controls (1,-2) .. (c);
    \node at (0,-2) {$-A_1$};
    \end{tikzpicture}\;.
    \label{eq:MergingContractedLoops}
\end{equation}
The decoration of the bottom quiver in \eqref{eq:MergingContractedLoops} is redundant and is only there in order to help illustrate the transition. After merging the decorated $\mathrm{U}(1)$s we get the expected quiver of \eqref{eq:MinB3}.

Returning to our quiver in \eqref{eq:DecorateIsDiscreteGauge} we can propose the `subtraction':
\begin{equation}
    \begin{tikzpicture}
        \node (a) at (0,0) {
    $\begin{tikzpicture}
        \node[gauge,label=below:{$1$}] (1) at (-1,0) {};
        \node[gauge,label=below:{$1$}] (2) at (0,0) {};
        \draw (1)--(2);
        \begin{scope}[rotate around={60:(-1,0)}]
        \node[gauge,label=below:{$1$}] (5) at (0,0) {};
        \end{scope}
        \draw (5)--(1);
        \draw[purple] (5) circle (0.6cm);
        \draw[purple] (2) circle (0.6cm);
    \end{tikzpicture}$};
        \node (c) at (0,-4) {
    $\mathcal{C}\left(\vcenter{\hbox{\scalebox{1}{\begin{tikzpicture}
        \node[gauge,label=below:{$1$}] (1) at (1,0) {};
        \node[gauge,label=below:{$1$}] (2) at (2,0) {};
        \draw[transform canvas={yshift=1pt}] (1)--(2);
        \draw[transform canvas={yshift=-1pt}] (1)--(2);
        \draw (1.5-0.1,0.2)--(1.5+0.1,0)--(1.5-0.1,-0.2);
        \draw[purple] (2) circle (0.6cm);
    \end{tikzpicture}}}}\right)=\mathbb{H}$};
    \draw[->] (a) .. controls (1,-2) .. (c);
    \node at (0,-2) {$-A_1$};
    \end{tikzpicture}\;.
    \label{eq:MergingContractedLoopsA1}
\end{equation}
Indeed, after an $A_1$ transition we end up with a quiver whose Coulomb branch is $\mathbb{H}$, as expected.\\

Let us now consider a more complicated example \cite{2015arXiv150205770F,Hanany:2018dvd,Bourget:2020bxh}:
\begin{equation}
    \mathcal{C}\left(\vcenter{\hbox{\scalebox{1}{
    \begin{tikzpicture}
        \node[gauge,label=below:{$1$}] (1) at (0,0) {};
        \node[gauge,label=below:{$2$}] (2) at (1,0) {};
        \node[gauge,label=below:{$3$}] (4) at (2,0) {};
        \draw (1)--(2)--(4);
        \draw (4) to [out=315,in=45,looseness=10] (4);
    \end{tikzpicture}}}}
    \right)=\overline{\textnormal{n.n.min G}{}_2}\qquad\textnormal{with Hasse diagram}\qquad
    \vcenter{\hbox{\scalebox{1}{
    \begin{tikzpicture}
        \node[hasse] (1) at (0,0) {};
        \node[hasse] (2) at (0,1) {};
        \node[hasse] (3) at (0,2) {};
        \node[hasse] (4) at (0,3) {};
        \draw (1)--(2)--(3)--(4);
        \node at (-0.4,0.5) {$g_2$};
        \node at (-0.4,1.5) {$m$};
        \node at (-0.4,2.5) {$A_1$};
    \end{tikzpicture}}}}\;.
    \label{eq:NNMinG2}
\end{equation}
The quiver for the bottom slice is
\begin{equation}
    \mathcal{C}\left(\vcenter{\hbox{\scalebox{1}{
    \begin{tikzpicture}
        \node[gauge,label=below:{$1$}] (1) at (0,0) {};
        \node[gauge,label=below:{$2$}] (2) at (1,0) {};
        \node[gauge,label=below:{$1$}] (4) at (2,0) {};
        \draw (1)--(2);
        \draw[transform canvas={yshift=2pt}] (2)--(4);
        \draw (2)--(4);
        \draw[transform canvas={yshift=-2pt}] (2)--(4);
        \draw (1.5-0.1,0.2)--(1.5+0.1,0)--(1.5-0.1,-0.2);
    \end{tikzpicture}}}}
    \right)=\overline{\textnormal{min G}{}_2}\qquad\textnormal{with Hasse diagram}\qquad
    \vcenter{\hbox{\scalebox{1}{
    \begin{tikzpicture}
        \node[hasse] (1) at (0,0) {};
        \node[hasse] (2) at (0,1) {};
        \draw (1)--(2);
        \node at (-0.4,0.5) {$g_2$};
    \end{tikzpicture}}}}\;.
    \label{eq:MinG2}
\end{equation}
We may replace the $\mathrm{U}(3)$ node with an adjoint in \eqref{eq:NNMinG2} with three $\mathrm{U}(1)$s decorated with the same colour:
\begin{equation}
    \vcenter{\hbox{\scalebox{1}{
    \begin{tikzpicture}
        \node[gauge,label=below:{$1$}] (1) at (0,0) {};
        \node[gauge,label=below:{$2$}] (2) at (1,0) {};
        \node[gauge,label=below:{$3$}] (4) at (2,0) {};
        \draw (1)--(2)--(4);
        \draw (4) to [out=315,in=45,looseness=10] (4);
    \end{tikzpicture}}}}=
    \vcenter{\hbox{\scalebox{1}{
    \begin{tikzpicture}
        \node[gauge,label=below:{$1$}] (1) at (0,0) {};
        \node[gauge,label=below:{$2$}] (2) at (1,0) {};
        \node[gauge,label=right:{$1$}] (4u) at (2,1) {};
        \node[gauge,label=right:{$1$}] (4) at (2,0) {};
        \node[gauge,label=right:{$1$}] (4d) at (2,-1) {};
        \draw (1)--(2)--(4) (4u)--(2)--(4d);
        \draw[purple] (4u) circle (0.5cm);
        \draw[purple] (4) circle (0.5cm);
        \draw[purple] (4d) circle (0.5cm);
    \end{tikzpicture}}}}\;.
    \label{eq:NNMinG2Decorated}
\end{equation}
Following what we did before, we can now contract two decorated $\mathrm{U}(1)$s into a single short $\mathrm{U}(1)$ realising an $A_1$ transition. We obtain a quiver with two decorated $\mathrm{U}(1)$s, one short, one long. In order to obtain the quiver in \eqref{eq:MinG2} we should now contract the decorated short $\mathrm{U}(1)$ with the decorated long $\mathrm{U}(1)$ to a single $\mathrm{U}(1)$ of shortness $1/3$, realising an $m$ transition:
\begin{equation}
    \begin{tikzpicture}
        \node (a) at (0,0) {
    $\begin{tikzpicture}
        \node[gauge,label=below:{$1$}] (1) at (0,0) {};
        \node[gauge,label=below:{$2$}] (2) at (1,0) {};
        \node[gauge,label=right:{$1$}] (4u) at (2,1) {};
        \node[gauge,label=right:{$1$}] (4) at (2,0) {};
        \node[gauge,label=right:{$1$}] (4d) at (2,-1) {};
        \draw (1)--(2)--(4) (4u)--(2)--(4d);
        \draw[purple] (4u) circle (0.5cm);
        \draw[purple] (4) circle (0.5cm);
        \draw[purple] (4d) circle (0.5cm);
    \end{tikzpicture}$};
        \node (c) at (0,-4) {
    $\begin{tikzpicture}
        \node[gauge,label=below:{$1$}] (1) at (0,0) {};
        \node[gauge,label=below:{$2$}] (2) at (1,0) {};
        \node[gauge,label=right:{$1$}] (4u) at (2,1) {};
        \node[gauge,label=right:{$1$}] (4) at (2,0) {};
        \draw (1)--(2)--(4u);
        \draw[transform canvas={yshift=1pt}] (2)--(4);
        \draw[transform canvas={yshift=-1pt}] (2)--(4);
        \draw (1.5-0.1,0.2)--(1.5+0.1,0)--(1.5-0.1,-0.2);
        \draw[purple] (4u) circle (0.5cm);
        \draw[purple] (4) circle (0.5cm);
    \end{tikzpicture}$};
    \draw[->] (a) .. controls (1,-2) .. (c);
    \node at (0,-2) {$-A_1$};
        \node (e) at (0,-8) {
    $\begin{tikzpicture}
        \node[gauge,label=below:{$1$}] (1) at (0,0) {};
        \node[gauge,label=below:{$2$}] (2) at (1,0) {};
        \node[gauge,label=below:{$1$}] (4) at (2,0) {};
        \draw (1)--(2)--(4);
        \draw[transform canvas={yshift=2pt}] (2)--(4);
        \draw[transform canvas={yshift=-2pt}] (2)--(4);
        \draw (1.5-0.1,0.2)--(1.5+0.1,0)--(1.5-0.1,-0.2);
        \draw[purple] (4) circle (0.5cm);
    \end{tikzpicture}$};
    \draw[->] (c) .. controls (1,-6) .. (e);
    \node at (0,-6) {$-m$};
    \end{tikzpicture}
    \label{eq:MergingContractedLoopsG2}
\end{equation}
The second quiver in \eqref{eq:MergingContractedLoopsG2} was discussed in \cite{Bourget:2020bxh}, albeit without an explicit mention of decoration, where it was concluded that its Coulomb branch is $\overline{\textnormal{n.min}G_2}$. This implies that our quiver subtractions are consistent with the Hasse diagram in \eqref{eq:NNMinG2}.\\

We can now generalise this algorithm to a rank $k$ unitary node with an adjoint.

\paragraph{Hasse diagram of $\mathrm{Sym}^k(\mathbb{C}^2)$.} We have
\begin{equation}
    \mathrm{Sym}^k(\mathbb{C}^2)=\mathcal{C}\left(\vcenter{\hbox{\scalebox{1}{
    \begin{tikzpicture}
        \node[gauge,label=below:{$1$}] (1) at (0,0) {};
        \node[gauge,label=left:{$k$}] (2) at (0,1) {};
        \draw (2) to [out=45,in=135,looseness=10] (2);
        \draw (1)--(2);
    \end{tikzpicture}}}}\right)
\end{equation}
With
\begin{equation}
    \vcenter{\hbox{\scalebox{1}{
    \begin{tikzpicture}
        \node[gauge,label=below:{$1$}] (1) at (0,0) {};
        \node[gauge,label=left:{$k$}] (2) at (0,1) {};
        \draw (2) to [out=45,in=135,looseness=10] (2);
        \draw (1)--(2);
    \end{tikzpicture}}}}=
    \vcenter{\hbox{\scalebox{1}{
    \begin{tikzpicture}
        \node[gauge,label=below:{$1$}] (1) at (0,0) {};
        \node[gauge,label=left:{$1$}] (2) at (-1,1) {};
        \node[gauge,label=right:{$1$}] (3) at (1,1) {};
        \node at (0,1) {$\cdots$};
        \draw [decorate,decoration={brace,amplitude=5pt}] (-1.2,1.2)--(1.2,1.2);
        \node at (0,1.6) {$k$};
        \draw[purple] (2) circle (0.5cm);
        \draw[purple] (3) circle (0.5cm);
        \draw (1)--(2) (1)--(3);
    \end{tikzpicture}}}}
\end{equation}
we can use our method of merging decorated $\mathrm{U}(1)$s in order to compute the Hasse diagram of $\mathrm{Sym}^k(\mathbb{C}^2)$.

A generic transition will be of the form
\begin{equation}
    \begin{tikzpicture}
        \node (a) at (0,0) {
    $\begin{tikzpicture}
        \node[gauge,label=below:{$1$}] (1) at (0,0) {};
        \node[gauge,label=left:{$1$}] (2) at (-0.5,1) {};
        \node[gauge,label=right:{$1$}] (3) at (0.5,1) {};
        \draw[thick,double] (1)--(2);
        \draw[thick,double] (1)--(3);
        \draw (-0.25+0.2,0.5)--(-0.25,0.5)--(-0.25-0.1,0.5-0.2);
        \draw (0.25-0.2,0.5)--(0.25,0.5)--(0.25+0.1,0.5-0.2);
        \node at (-0.5,0.5) {$a$};
        \node at (0.5,0.5) {$b$};
        \draw[purple] (2) circle (0.5cm);
        \draw[purple] (3) circle (0.5cm);
    \end{tikzpicture}$};
        \node (c) at (0,-4) {
    $\begin{tikzpicture}
        \node[gauge,label=below:{$1$}] (1) at (0,0) {};
        \node[gauge,label=left:{$1$}] (2) at (0,1) {};
        \draw[thick,double] (1)--(2);
        \draw (-0.2,0.5-0.2)--(0,0.5)--(0+0.2,0.5-0.2);
        \node at (-0.7,0.5) {$a+b$};
        \draw[purple] (2) circle (0.5cm);
    \end{tikzpicture}$};
    \draw[->] (a) .. controls (1,-2) .. (c);
    \node at (0,-2) {$-K_{a,b}$};
    \end{tikzpicture}\;.
\end{equation}

\paragraph{New slices from merging decorated $\mathrm{U}(1)$s.}
We can now conjecture a family of slices indexed by two numbers $a$ and $b$
\begin{equation}
K_{a,b}\times\mathbb{H}=\mathcal{C}\left(\vcenter{\hbox{\scalebox{1}{
    \begin{tikzpicture}
        \node[gauge,label=below:{$1$}] (1) at (0,0) {};
        \node[gauge,label=left:{$1$}] (2) at (-0.5,1) {};
        \node[gauge,label=right:{$1$}] (3) at (0.5,1) {};
        \draw[thick,double] (1)--(2);
        \draw[thick,double] (1)--(3);
        \draw (-0.25+0.2,0.5)--(-0.25,0.5)--(-0.25-0.1,0.5-0.2);
        \draw (0.25-0.2,0.5)--(0.25,0.5)--(0.25+0.1,0.5-0.2);
        \node (a) at (-0.5,0.5) {$a$};
        \node (b) at (0.5,0.5) {$b$};
        \draw[purple] (3) circle (0.5cm);
        \draw[purple] (2) circle (0.5cm);
    \end{tikzpicture}}}}
    \right)
\end{equation}
At the moment we do not know how to compute this Coulomb branch directly, except for the case $a=b$ for which we get:
\begin{equation}
    K_{a,a}\times\mathbb{H}=\mathcal{C}\left(
    \vcenter{\hbox{\scalebox{1}{
    \begin{tikzpicture}
        \node[gauge,label=below:{$1$}] (1) at (0,0) {};
        \node[gauge,label=left:{$1$}] (2) at (-0.5,1) {};
        \node[gauge,label=right:{$1$}] (3) at (0.5,1) {};
        \draw[thick,double] (1)--(2);
        \draw[thick,double] (1)--(3);
        \draw (-0.25+0.2,0.5)--(-0.25,0.5)--(-0.25-0.1,0.5-0.2);
        \draw (0.25-0.2,0.5)--(0.25,0.5)--(0.25+0.1,0.5-0.2);
        \node (a) at (-0.5,0.5) {$a$};
        \node (b) at (0.5,0.5) {$a$};
        \draw[purple] (3) circle (0.5cm);
        \draw[purple] (2) circle (0.5cm);
    \end{tikzpicture}}}}
    \right)
    =\mathcal{C}\left(\vcenter{\hbox{\scalebox{1}{
    \begin{tikzpicture}
        \node[gauge,label=below:{$1$}] (1) at (0,0) {};
        \node[gauge,label=left:{$2$}] (2) at (0,1) {};
        \draw[thick,double] (1)--(2);
        \draw (-0.2,0.5-0.2)--(0,0.5)--(0+0.2,0.5-0.2);
        \node (a) at (-0.2,0.5) {$a$};
        \draw (2) to [out=45,in=135,looseness=10] (2);
    \end{tikzpicture}}}}
    \right)=A_1\times\mathbb{H}
\end{equation}

Comparing with the explicit computations of slices in $\mathrm{Sym}^k(\mathbb{C}^2)$ in Appendix \ref{app:SYM} we find that
\begin{equation}
    K_{2,1}=K_{3,1}=m\;.
\end{equation}
We conjecture that
\begin{equation}
    K_{a,b}=\left\{\begin{matrix}
    A_1 & \textnormal{if }a=b\\
    m & \textnormal{if }a\neq b
    \end{matrix}
    \right.\;.
\end{equation}
At the moment no definition of the Coulomb branch (or at least its Hilbert series) of a decorated quiver exists. Finding such a definition is necessary to understand all slices in the moduli space of instantons from a Coulomb branch perspective.

With this we are ready to face the moduli space of instantons using quiver subtraction on CFHM quivers.

\paragraph{Hasse diagram of $\mathcal{M}_{4,\mathrm{SU}(3)}$.} We present the Hasse diagram of the moduli space of $4$ $\mathrm{SU}(3)$ instantons obtained from quiver subtraction in Figure \ref{fig:QuiverSubtractionHasse}.

\begin{landscape}
\begin{figure}
    \centering
    \vspace*{-2cm}
             \begin{tikzpicture}
                \node (4) at (0,0) {
    $\begin{tikzpicture}
        \node[gauge,label=below:{$1$}] (a) at (-1,0) {};
        \node[gauge,label=left:{$1$}] (u) at (-1,1) {};
        \draw[transform canvas={xshift=3pt}] (a)--(u);
        \draw[transform canvas={xshift=1pt}] (a)--(u);
        \draw[transform canvas={xshift=-1pt}] (a)--(u);
        \draw[transform canvas={xshift=-3pt}] (a)--(u);
        \draw (-1-0.2,0.5-0.1)--(-1,0.5+0.1)--(-1+0.2,0.5-0.1);
    \end{tikzpicture}$};
                \node (22) at (-1.5,3.5) {
    $\begin{tikzpicture}
        \node[gauge,label=below:{$1$}] (a) at (-1,0) {};
        \begin{scope}[rotate around={45:(-1,0)}]
        \node[gauge,label=left:{$1$}] (u) at (-1,1) {};
        \draw[transform canvas={xshift=1.3pt,yshift=1.3pt}] (a)--(u);
        \draw[transform canvas={xshift=-1.3pt,yshift=-1.3pt}] (a)--(u);
        \draw (-1-0.2,0.5-0.1)--(-1,0.5+0.1)--(-1+0.2,0.5-0.1);
        \end{scope}
        \begin{scope}[rotate around={-45:(-1,0)}]
        \node[gauge,label=left:{$1$}] (u2) at (-1,1) {};
        \draw[transform canvas={xshift=1.3pt,yshift=-1.3pt}] (a)--(u2);
        \draw[transform canvas={xshift=-1.3pt,yshift=1.3pt}] (a)--(u2);
        \draw (-1-0.2,0.5-0.1)--(-1,0.5+0.1)--(-1+0.2,0.5-0.1);
        \end{scope}
        \draw[purple] (u) circle (0.6cm);
        \draw[purple] (u2) circle (0.6cm);
    \end{tikzpicture}$};
                \node (31) at (1.5,3.5) {
    $\begin{tikzpicture}
        \node[gauge,label=below:{$1$}] (a) at (-1,0) {};
        \begin{scope}[rotate around={45:(-1,0)}]
        \node[gauge,label=left:{$1$}] (u) at (-1,1) {};
        \draw[transform canvas={xshift=1.5pt,yshift=1.5pt}] (a)--(u);
        \draw (a)--(u);
        \draw[transform canvas={xshift=-1.5pt,yshift=-1.5pt}] (a)--(u);
        \draw (-1-0.2,0.5-0.1)--(-1,0.5+0.1)--(-1+0.2,0.5-0.1);
        \end{scope}
        \begin{scope}[rotate around={-45:(-1,0)}]
        \node[gauge,label=left:{$1$}] (u2) at (-1,1) {};
        \draw (a)--(u2);
        \end{scope}
        \draw[purple] (u) circle (0.6cm);
        \draw[purple] (u2) circle (0.6cm);
    \end{tikzpicture}$};
                \node (211) at (0,7) {
    $\begin{tikzpicture}
        \node[gauge,label=below:{$1$}] (a) at (-1,0) {};
        \begin{scope}[rotate around={45:(-1,0)}]
        \node[gauge,label=left:{$1$}] (u) at (-1,1) {};
        \draw[transform canvas={xshift=1.3pt,yshift=1.3pt}] (a)--(u);
        \draw[transform canvas={xshift=-1.3pt,yshift=-1.3pt}] (a)--(u);
        \draw (-1-0.2,0.5-0.1)--(-1,0.5+0.1)--(-1+0.2,0.5-0.1);
        \end{scope}
        \begin{scope}[rotate around={-45:(-1,0)}]
        \node[gauge,label=left:{$2$}] (u2) at (-1,1) {};
        \draw (a)--(u2);
        \end{scope}
        \draw[purple] (u) circle (0.6cm);
        \draw[purple] (u2) circle (0.6cm);
        \draw (u2) to [out=45,in=135,looseness=8] (u2);
    \end{tikzpicture}$};
                \node (1111) at (0,10.5) {
    $\begin{tikzpicture}
        \node[gauge,label=below:{$1$}] (a) at (-1,0) {};
        \node[gauge,label=left:{$4$}] (u) at (-1,1) {};
        \draw (a)--(u);
        \draw (u) to [out=45,in=135,looseness=8] (u);
    \end{tikzpicture}$};
                \draw (4)--(31)--(211)--(1111) (4)--(22)--(211);
                \node (3) at (5,4.5) {
    $\begin{tikzpicture}
        \node[gauge,label=below:{$1$}] (a) at (-1,0) {};
        \node[gauge,label=below:{$1$}] (b) at (0,0) {};
        \node[gauge,label=right:{$1$}] (c) at (1,0.5) {};
        \node[gauge,label=right:{$1$}] (d) at (1,-0.5) {};
        \node[gauge,label=left:{$1$}] (u) at (-1,1) {};
        \draw (a)--(b)--(c)--(d)--(b);
        \draw[transform canvas={xshift=3pt}] (a)--(u);
        \draw (a)--(u);
        \draw[transform canvas={xshift=-3pt}] (a)--(u);
        \draw (-1-0.2,0.5-0.1)--(-1,0.5+0.1)--(-1+0.2,0.5-0.1);
        \draw[purple] (u) circle (0.6cm);
        \draw[purple] \convexpath{b,c,d}{0.6cm};
    \end{tikzpicture}$};
                \node (21) at (5,8) {
    $\begin{tikzpicture}
        \node[gauge,label=below:{$1$}] (a) at (-1,0) {};
        \node[gauge,label=below:{$1$}] (b) at (0,0) {};
        \node[gauge,label=right:{$1$}] (c) at (1,0.5) {};
        \node[gauge,label=right:{$1$}] (d) at (1,-0.5) {};
        \draw (a)--(b)--(c)--(d)--(b);
        \begin{scope}[rotate around={45:(-1,0)}]
        \node[gauge,label=left:{$1$}] (u) at (-1,1) {};
        \draw[transform canvas={xshift=1.3pt,yshift=1.3pt}] (a)--(u);
        \draw[transform canvas={xshift=-1.3pt,yshift=-1.3pt}] (a)--(u);
        \draw (-1-0.2,0.5-0.1)--(-1,0.5+0.1)--(-1+0.2,0.5-0.1);
        \end{scope}
        \begin{scope}[rotate around={-45:(-1,0)}]
        \node[gauge,label=left:{$1$}] (u2) at (-1,1) {};
        \draw (a)--(u2);
        \end{scope}
        \draw[purple] (u) circle (0.6cm);
        \draw[purple] (u2) circle (0.6cm);
        \draw[purple] \convexpath{b,c,d}{0.6cm};
    \end{tikzpicture}$};
                \node (111) at (5,11.5) {
    $\begin{tikzpicture}
        \node[gauge,label=below:{$1$}] (a) at (-1,0) {};
        \node[gauge,label=below:{$1$}] (b) at (0,0) {};
        \node[gauge,label=right:{$1$}] (c) at (1,0.5) {};
        \node[gauge,label=right:{$1$}] (d) at (1,-0.5) {};
        \node[gauge,label=left:{$3$}] (u) at (-1,1) {};
        \draw (a)--(b)--(c)--(d)--(b);
        \draw (a)--(u);
        \draw (u) to [out=45,in=135,looseness=8] (u);
        \draw[purple] (u) circle (0.6cm);
        \draw[purple] \convexpath{b,c,d}{0.6cm};
    \end{tikzpicture}$};
                \draw (3)--(21)--(111);
                \node (2) at (10,9) {
    $\begin{tikzpicture}
        \node[gauge,label=below:{$1$}] (a) at (-1,0) {};
        \node[gauge,label=below:{$2$}] (b) at (0,0) {};
        \node[gauge,label=right:{$2$}] (c) at (1,0.5) {};
        \node[gauge,label=right:{$2$}] (d) at (1,-0.5) {};
        \node[gauge,label=left:{$1$}] (u) at (-1,1) {};
        \draw (a)--(b)--(c)--(d)--(b);
        \draw[transform canvas={xshift=2pt}] (a)--(u);
        \draw[transform canvas={xshift=-2pt}] (a)--(u);
        \draw (-1-0.2,0.5-0.1)--(-1,0.5+0.1)--(-1+0.2,0.5-0.1);
        \draw[purple] (u) circle (0.6cm);
        \draw[purple] \convexpath{b,c,d}{0.6cm};
    \end{tikzpicture}$};
                \node (11) at (10,12.5) {
    $\begin{tikzpicture}
        \node[gauge,label=below:{$1$}] (a) at (-1,0) {};
        \node[gauge,label=below:{$2$}] (b) at (0,0) {};
        \node[gauge,label=right:{$2$}] (c) at (1,0.5) {};
        \node[gauge,label=right:{$2$}] (d) at (1,-0.5) {};
        \node[gauge,label=left:{$2$}] (u) at (-1,1) {};
        \draw (a)--(b)--(c)--(d)--(b);
        \draw (a)--(u);
        \draw (u) to [out=45,in=135,looseness=8] (u);
        \draw[purple] (u) circle (0.6cm);
        \draw[purple] \convexpath{b,c,d}{0.6cm};
    \end{tikzpicture}$};
                \draw (2)--(11);
                \node (1) at (15,13.5) {
    $\begin{tikzpicture}
        \node[gauge,label=below:{$1$}] (a) at (-1,0) {};
        \node[gauge,label=below:{$3$}] (b) at (0,0) {};
        \node[gauge,label=right:{$3$}] (c) at (1,0.5) {};
        \node[gauge,label=right:{$3$}] (d) at (1,-0.5) {};
        \node[gauge,label=left:{$1$}] (u) at (-1,1) {};
        \draw (a)--(b)--(c)--(d)--(b);
        \draw (a)--(u);
        \draw[purple] (u) circle (0.6cm);
        \draw[purple] \convexpath{b,c,d}{0.6cm};
    \end{tikzpicture}$};
                \node (0) at (20,14.5) {
    $\begin{tikzpicture}
        \node[gauge,label=below:{$1$}] (a) at (-1,0) {};
        \node[gauge,label=below:{$4$}] (b) at (0,0) {};
        \node[gauge,label=right:{$4$}] (c) at (1,0.5) {};
        \node[gauge,label=right:{$4$}] (d) at (1,-0.5) {};
        \draw (a)--(b)--(c)--(d)--(b);
    \end{tikzpicture}$};
                \draw[red] (31)--(3) (211)--(21)--(2) (1111)--(111)--(11)--(1)--(0);
                \node[label=below:{$a_2$}] at ($(1)!0.5!(0)$) {};
                \node[label=below:{$a_2$}] at ($(11)!0.5!(1)$) {};
                \node[label=below:{$a_2$}] at ($(111)!0.5!(11)$) {};
                \node[label=below:{$a_2$}] at ($(1111)!0.5!(111)$) {};
                \node[label=below:{$a_2$}] at ($(21)!0.5!(2)$) {};
                \node[label=below:{$a_2$}] at ($(211)!0.5!(21)$) {};
                \node[label=below:{$a_2$}] at ($(31)!0.5!(3)$) {};
                \node[label=right:{$A_1$}] at ($(2)!0.5!(11)$) {};
                \node[label=right:{$A_1$}] at ($(21)!0.5!(111)$) {};
                \node[label=left:{$A_1$}] at ($(211)!0.5!(22)$) {};
                \node[label=right:{$A_1$}] at ($(211)!0.5!(1111)$) {};
                \node[label=left:{$A_1$}] at ($(4)!0.5!(22)$) {};
                \node[label=right:{$m$}] at ($(3)!0.5!(21)$) {};
                \node[label=right:{$m$}] at ($(31)!0.5!(211)$) {};
                \node[label=right:{$m$}] at ($(4)!0.5!(31)$) {};
            \end{tikzpicture}
    \caption{Quiver subtraction algorithm for the moduli space of $4$ $\mathrm{SU}(3)$ instantons. Decoration is only added where needed.}
    \label{fig:QuiverSubtractionHasse}
\end{figure}

\end{landscape}

\subsection{Brane Webs}

Let us now turn to brane systems, and try to realise the quiver subtractions of the previous section in a brane web.

There are several more brane systems, other than branes within branes, which realise instanton moduli spaces. In \cite{Cremonesi:2014xha} a system of D3, D5 and NS5 branes on a circle was studied. For us it is more convenient to realise the moduli space of instantons as the Higgs branch of a $5d$ $\mathcal{N}=1$ theory living on a $5$-brane web suspended between $7$-branes, avoiding the complications which arise due a circular direction. We focus on the particular example of $\mathrm{SU}(3)$ instantons. The moduli space of $k$ $\mathrm{SU}(3)$ instantons is realised as one cone in the Higgs branch of the SCFT living on the brane web in \eqref{eq:webSU3}. It is the cone corresponding to a particular subweb decomposition, whose magnetic quiver $\mathsf{Q}_{k,\mathrm{SU}(3)}$ can be obtained from the rules given in \cite{Cabrera:2018jxt}.
\begin{equation}
    \vcenter{\hbox{\scalebox{1}{\begin{tikzpicture}
        \node[seven] (0) at (0,0) {};
        \node[seven] (l) at (1,0) {};
        \node[seven] (lu) at (1,1) {};
        \node[seven] (u) at (2,1) {};
        \node[seven] (d) at (2,-1) {};
        \node[seven] (rd) at (3,-1) {};
        \node[seven] (r) at (3,0) {};
        \draw (0)--(l);
        \draw[double,orange] (l)--(r);
        \draw[double,red] (u)--(d);
        \draw[double,blue] (lu)--(rd);
        \node at (1.4,-0.2) {$k$};
        \node at (2.2,0.5) {$k$};
        \node at (2.7,-0.4) {$k$};
    \end{tikzpicture}}}}
    \qquad\mathsf{Q}_{k,\mathrm{SU}(3)}=
    \vcenter{\hbox{\scalebox{1}{\begin{tikzpicture}
        \node[gauge,label=below:{$1$}] (1) at (0,0) {};
        \node[gaugeo,label=below:{$k$}] (2) at (1,0) {};
        \node[gauger,label=right:{$k$}] (3) at (2,0.5) {};
        \node[gaugeb,label=right:{$k$}] (4) at (2,-0.5) {};
        \draw (1)--(2)--(3)--(4)--(2);
    \end{tikzpicture}}}}
    \label{eq:webSU3}
\end{equation}
For concreteness, let us consider $k=2$. On a general point on the moduli space all of the subwebs are separated. In order to move to a lower leaf, we have to align the orange, red and blue coloured branes in \eqref{eq:webSU3}. Since there are two of each available, we can do this twice, corresponding to two transitions. Since we encounter subwebs with self intersection zero, adjoint hypermultiplets make an appearance in the magnetic quiver, according to \cite[Appendix B]{Bourget:2020mez}. Furthermore the magnetic quiver should be decorated according to \cite[Appendix C]{Bourget:2020mez}. We provide the magnetic quiver for each leaf:
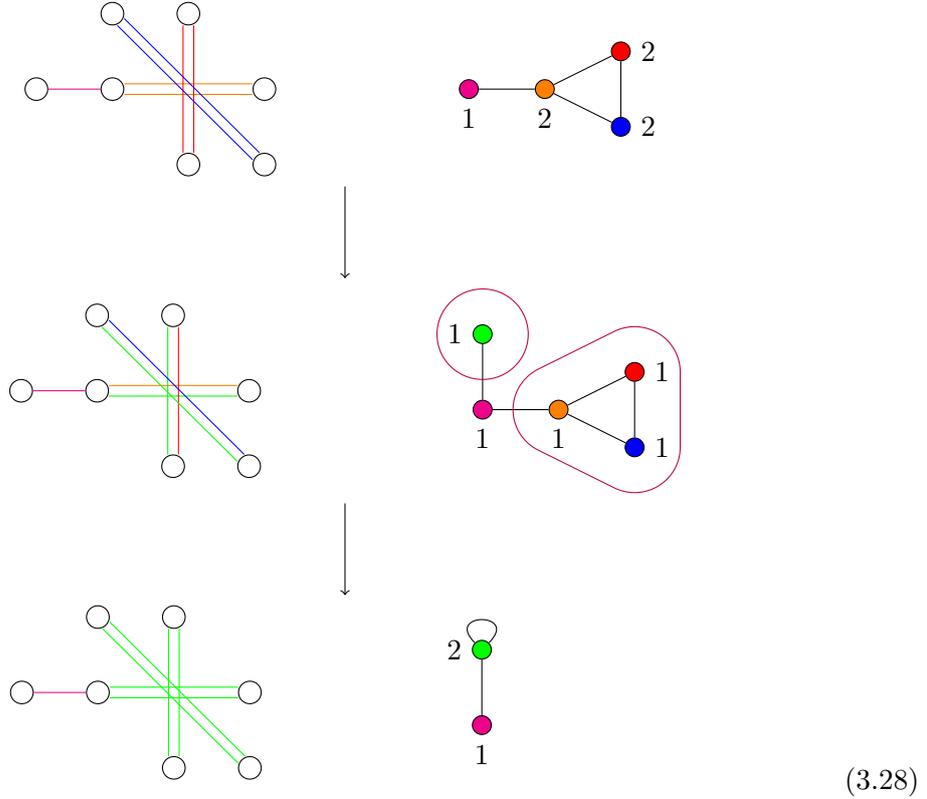
\begin{equation}
    \begin{tikzpicture}
        \node (top) at (0,0) {$\vcenter{\hbox{\scalebox{1}{\begin{tikzpicture}
        \node[seven] (0) at (0,0) {};
        \node[seven] (l) at (1,0) {};
        \node[seven] (lu) at (1,1) {};
        \node[seven] (u) at (2,1) {};
        \node[seven] (d) at (2,-1) {};
        \node[seven] (rd) at (3,-1) {};
        \node[seven] (r) at (3,0) {};
        \draw[magenta] (0)--(l);
        \draw[orange,transform canvas={yshift=2pt}] (l)--(r);
        \draw[orange,transform canvas={yshift=-2pt}] (l)--(r);
        \draw[red,transform canvas={xshift=2pt}] (u)--(d);
        \draw[red,transform canvas={xshift=-2pt}] (u)--(d);
        \draw[blue,transform canvas={xshift=1.3pt,yshift=1.3pt}] (lu)--(rd);
        \draw[blue,transform canvas={xshift=-1.3pt,yshift=-1.3pt}] (lu)--(rd);
    \end{tikzpicture}}}}
    \qquad\qquad\qquad
    \vcenter{\hbox{\scalebox{1}{\begin{tikzpicture}
        \node[gaugem,label=below:{$1$}] (1) at (0,0) {};
        \node[gaugeo,label=below:{$2$}] (2) at (1,0) {};
        \node[gauger,label=right:{$2$}] (3) at (2,0.5) {};
        \node[gaugeb,label=right:{$2$}] (4) at (2,-0.5) {};
        \draw (1)--(2)--(3)--(4)--(2);
    \end{tikzpicture}}}}$};
    \node (mid) at (0,-4) {$\vcenter{\hbox{\scalebox{1}{\begin{tikzpicture}
        \node[seven] (0) at (0,0) {};
        \node[seven] (l) at (1,0) {};
        \node[seven] (lu) at (1,1) {};
        \node[seven] (u) at (2,1) {};
        \node[seven] (d) at (2,-1) {};
        \node[seven] (rd) at (3,-1) {};
        \node[seven] (r) at (3,0) {};
        \draw[magenta] (0)--(l);
        \draw[orange,transform canvas={yshift=2pt}] (l)--(r);
        \draw[green,transform canvas={yshift=-2pt}] (l)--(r);
        \draw[red,transform canvas={xshift=2pt}] (u)--(d);
        \draw[green,transform canvas={xshift=-2pt}] (u)--(d);
        \draw[blue,transform canvas={xshift=1.3pt,yshift=1.3pt}] (lu)--(rd);
        \draw[green,transform canvas={xshift=-1.3pt,yshift=-1.3pt}] (lu)--(rd);
    \end{tikzpicture}}}}
    \qquad\qquad\qquad
    \vcenter{\hbox{\scalebox{1}{\begin{tikzpicture}
        \node[gaugem,label=below:{$1$}] (1) at (0,0) {};
        \node[gaugeo,label=below:{$1$}] (2) at (1,0) {};
        \node[gauger,label=right:{$1$}] (3) at (2,0.5) {};
        \node[gaugeb,label=right:{$1$}] (4) at (2,-0.5) {};
        \node[gaugeg,label=left:{$1$}] (u) at (0,1) {};
        \draw (1)--(2)--(3)--(4)--(2);
        \draw (1)--(u);
        \draw[purple] (u) circle (0.6cm);
        \draw[purple] \convexpath{2,3,4}{0.6cm};
    \end{tikzpicture}}}}$};
    \node (bot) at (0,-8) {$\vcenter{\hbox{\scalebox{1}{\begin{tikzpicture}
        \node[seven] (0) at (0,0) {};
        \node[seven] (l) at (1,0) {};
        \node[seven] (lu) at (1,1) {};
        \node[seven] (u) at (2,1) {};
        \node[seven] (d) at (2,-1) {};
        \node[seven] (rd) at (3,-1) {};
        \node[seven] (r) at (3,0) {};
        \draw[magenta] (0)--(l);
        \draw[green,transform canvas={yshift=2pt}] (l)--(r);
        \draw[green,transform canvas={yshift=-2pt}] (l)--(r);
        \draw[green,transform canvas={xshift=2pt}] (u)--(d);
        \draw[green,transform canvas={xshift=-2pt}] (u)--(d);
        \draw[green,transform canvas={xshift=1.3pt,yshift=1.3pt}] (lu)--(rd);
        \draw[green,transform canvas={xshift=-1.3pt,yshift=-1.3pt}] (lu)--(rd);
    \end{tikzpicture}}}}
    \qquad\qquad\qquad
    \vcenter{\hbox{\scalebox{1}{\begin{tikzpicture}
        \node[gaugem,label=below:{$1$}] (1) at (0,0) {};
        \node[gaugeg,label=left:{$2$}] (u) at (0,1) {};
        \node (4) at (2,0.5) {};
        \draw[white] (4) circle (0.6cm);
        \draw (1)--(u);
        \draw (u) to [out=45,in=135,looseness=8] (u);
    \end{tikzpicture}}}}$};
    \draw[->] (top)--(mid);
    \draw[->] (mid)--(bot);
    \end{tikzpicture}
    \label{eq:braneWeb2Inst}
\end{equation}
We are left with a brane web decomposition described by the magnetic quiver $\mathrm{U}(2)$ with one adjoint and one fundamental hypermultiplet. The Coulomb branch of this magnetic quiver is $\mathbb{H}\times A_1$. Therefore we should expect another singular transition in the brane system.\\

In order to identify this transition it is instructive to first consider a simpler brane system.

\paragraph{Self-intersection zero and minus two.} The magnetic quivers obtained from the brane web decompositions have an interpretation as $3d$ theories. It is instructive to study D3-NS5 brane systems described by $3d$ theories at low energy, before we turn more complicated brane systems. In particular we consider
\begin{enumerate}
    \item[$(a)$] Two parallel D3 branes between two NS5 branes. The quiver description is
    \begin{equation}
        \begin{tikzpicture}
        \node[gauge,label=left:{$2$}] (u) at (0,1) {};
    \end{tikzpicture}
    \end{equation}
    \item[$(b)$] Two parallel D3 branes on a circle ending on a single NS5 brane. The quiver description is
    \begin{equation}
        \begin{tikzpicture}
        \node[gauge,label=left:{$2$}] (u) at (0,1) {};
        \draw (u) to [out=45,in=135,looseness=8] (u);
    \end{tikzpicture}
    \end{equation}
\end{enumerate}

\begin{figure}
    \centering
    \begin{tikzpicture}
    \node at (-9.5,5) {$(a)$};
    \node at (-1.5,6) {$(b)$};
        \node[hasse] (1) at (-3,0) {};
        \node[hasse] (2) at (-3,4) {};
        \node at (-3.5,2) {$A_1$};
        \node at (0,4) {\scalebox{.8}{\begin{tikzpicture}
\draw (8,-1)--(8,2);
\draw (7,0) arc[start angle=120,end angle=480,x radius=2cm,y radius =1cm];
\draw (7,.7) arc[start angle=120,end angle=480,x radius=2cm,y radius =1cm];
\node at (8.5,1.5) {NS5};
\node at (10.5,-.3) {D3};
\node at (10.5,-1) {D3};
        \end{tikzpicture}}};
        \node at (0,0) {\scalebox{.8}{\begin{tikzpicture}
\draw (8,-1)--(8,1);
\draw[very thick] (7,0) arc[start angle=120,end angle=480,x radius=2cm,y radius =1cm];
\node at (8.5,1.5) {NS5};
\node at (10.5,-1) {D3};
\node at (10,0) {\textbf{2}};
        \end{tikzpicture}}};
        \draw (1)--(2);
        \node[hasse] (3) at (-11,2) {};    
        \node at (-8,2) {\scalebox{.8}{\begin{tikzpicture}
\draw (0,-1)--(0,1.5);
\draw (2,-1)--(2,1.5);
\draw (0,0) arc[start angle=120,end angle=420,x radius=2cm,y radius =1cm];
\draw (0,.5) arc[start angle=120,end angle=420,x radius=2cm,y radius =1cm];
        \end{tikzpicture}}};   
        \end{tikzpicture}
    \caption{Brane systems for the two theories $(a)$ and $(b)$ along with the 3d $\mathcal{N}=4$ Coulomb branch Hasse diagrams. In case $(a)$ the Coulomb branch is smooth so the Hasse diagram is a point. In case $(b)$ there are two leaves. The bottom diagram shows the two coincident D3 branes. }
    \label{fig:TwoSystems}
\end{figure}
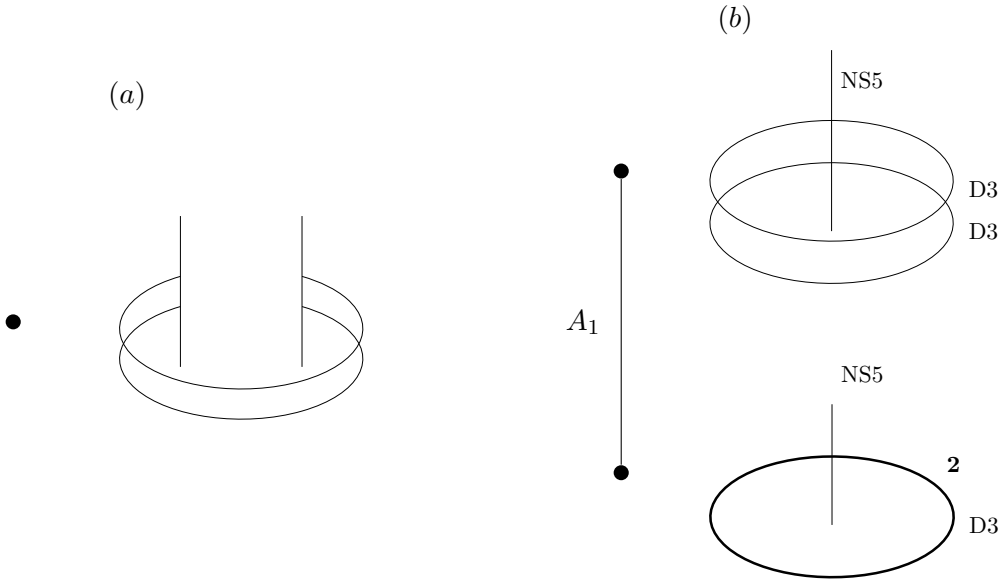

Both theories consist of a $\mathrm{U}(2)$ gauge node. In $(b)$ there is also an adjoint hypermultiplet which is not present in $(a)$. This adjoint hypermultiplet has a strong effect on the moduli spaces, whose Hasse diagrams are shown in Figure \ref{fig:TwoSystems}.

The moduli space of $(a)$ -- consisting of only the Coulomb branch -- is smooth \cite{atiyah2014geometry,Seiberg:1994aj,Chalmers:1996xh}. The gauge group is always $\mathrm{U}(1)^2$. The classical picture of branes coinciding leading to extra massless states breaks down at the quantum level.

The moduli space of $(b)$ -- which has enhanced $\mathcal{N}=8$ supersymmetry -- is not smooth: on a generic point the gauge group is broken to $\mathrm{U}(1)^2$, on a singular point it is enhanced to $\mathrm{U}(2)$. This corresponds to D3 branes becoming coincident. The latter effect can also be seen in the T-dual picture, which consists only of 2 parallel D2 branes.

\begin{figure}
    \centering
\begin{tikzpicture}
\node at (1,2) {$(a)$};
\node at (8,2) {$(b)$};
\draw (0,-1)--(0,1);
\draw (2,-1)--(2,1);
\draw (8,-1)--(8,1);
\node at (-1.8,-1) {D3};
\node at (5.2,-1) {D3};
\node at (0,1.5) {NS5};
\node at (2,1.5) {NS5};
\node at (8,1.5) {NS5};
\node at (-.5,.4) {\textcolor{red}{$-1$}};
\node at (2.5,.4) {\textcolor{red}{$-1$}};
\node at (7.5,.6) {\textcolor{red}{$-1$}};
\node at (8.5,.6) {\textcolor{red}{$-1$}};
\node at (7.5,-.3) {\textcolor{red}{$+1$}};
\node at (8.5,-.3) {\textcolor{red}{$+1$}};
\node at (1,-2.7) {$\mathrm{D3} \cdot \mathrm{D3}  = -2$};
\node at (8,-2.7) {$\mathrm{D3} \cdot \mathrm{D3} = 0$};
\draw (0,0) arc[start angle=120,end angle=420,x radius=2cm,y radius =1cm];
\draw (7,0) arc[start angle=120,end angle=480,x radius=2cm,y radius =1cm];
        \end{tikzpicture}
    \caption{Two configurations of D3 and NS5 branes on a circle. The self-intersection of the D3 brane is computed by adding the contributions in red.  }
    \label{fig:exampleSelfIntersection}
\end{figure}
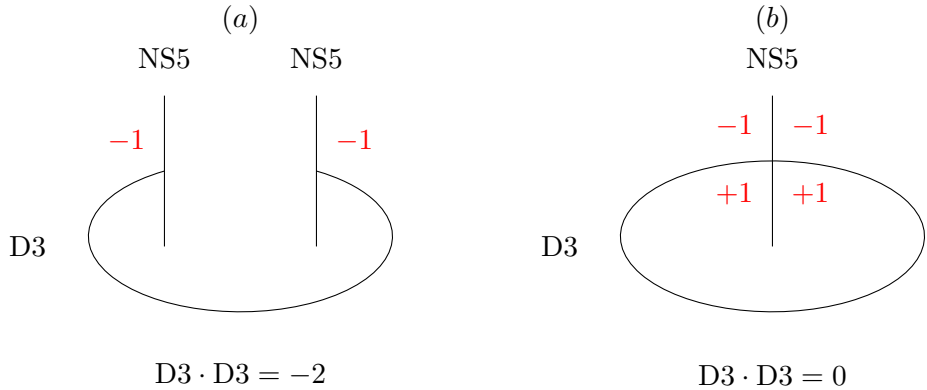

In both brane systems $(a)$ and $(b)$ we are dealing with parallel D3 branes. But only in one brane system there are extra massless states when the D3 branes coincide. The key difference between the two situations is the D3 self-intersection number.
The intersection number for two D3 branes is computed as follows: There is a $+1$ contribution if the D3 branes end on the same NS5 from opposite sides, and there is a $-1$ contribution if they end on the same NS5 from the same side. Note that this is a simpler version of the intersection number given for $5$-brane webs in Appendix \ref{app:braneweb} Equation \eqref{eq:IntersectionNumber}, as there is no notion of stable intersection for D3 branes.
The self-intersection number of a D3 brane is now just the intersection of a D3 brane with itself, see Figure \ref{fig:exampleSelfIntersection} for the computation to examples (a) and (b).

From this discussion it is suggestive to propose that \emph{coincident branes can lead to extra massless states when they have self-intersection $0$, and not when they have self-intersection $-2$}.

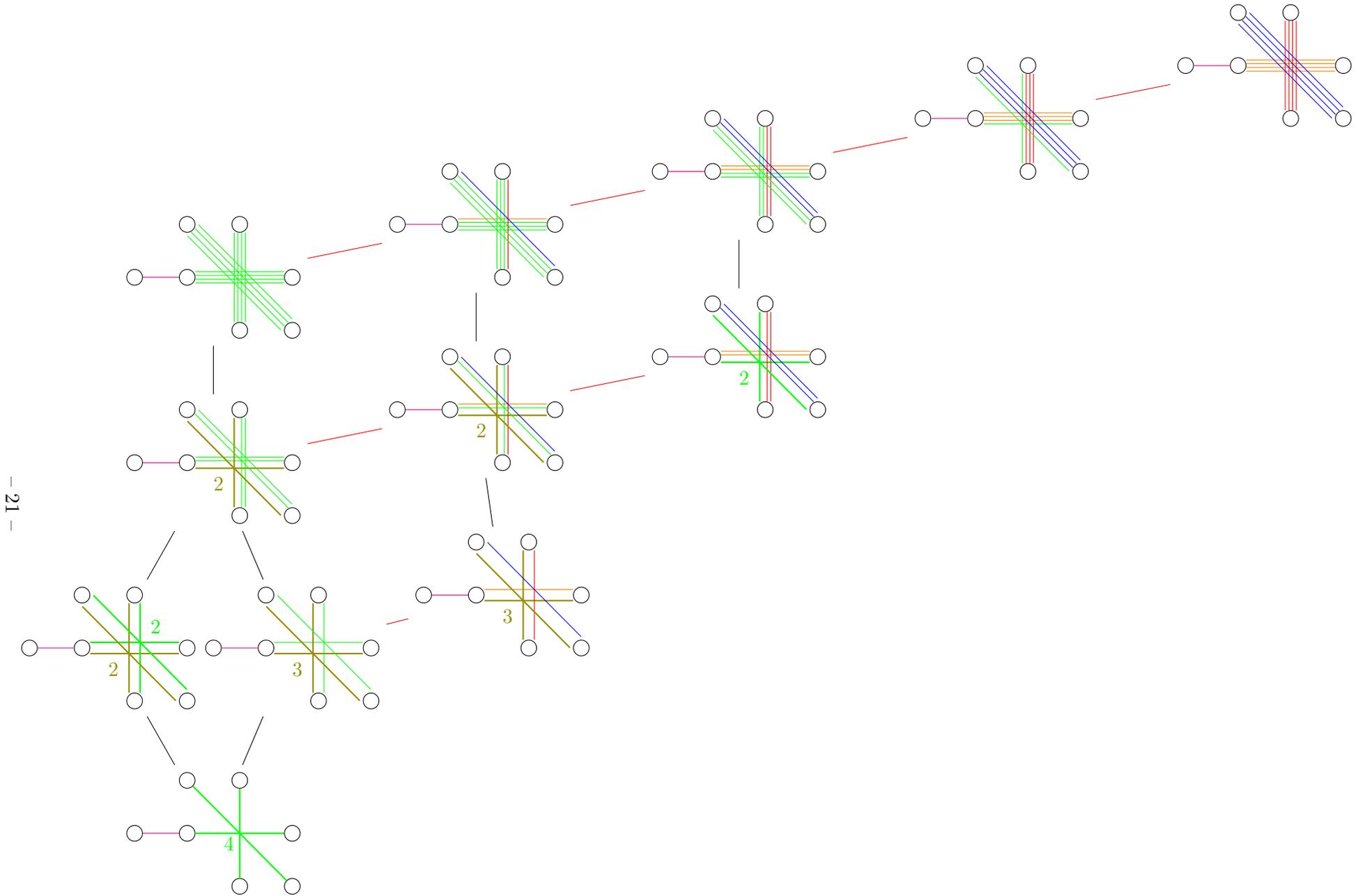
\begin{landscape}
\begin{figure}
    \centering
    \vspace*{-2cm}
    \hspace*{-2cm}
             \begin{tikzpicture}
                \node (4) at (0,0) {
    $\scalebox{1}{\begin{tikzpicture}
        \node[seven] (o) at (0,0) {};
        \node[seven] (l) at (1,0) {};
        \node[seven] (lu) at (1,1) {};
        \node[seven] (u) at (2,1) {};
        \node[seven] (d) at (2,-1) {};
        \node[seven] (rd) at (3,-1) {};
        \node[seven] (r) at (3,0) {};
        \draw[magenta] (o)--(l);
        \draw[green,thick] (l)--(r);
        \draw[green,thick] (u)--(d);
        \draw[green,thick] (lu)--(rd);
        \node at (1.8,-0.2) {\color{green}$4$};
    \end{tikzpicture}}$};
                \node (22) at (-2,3.5) {
    $\scalebox{1}{\begin{tikzpicture}
        \node[seven] (o) at (0,0) {};
        \node[seven] (l) at (1,0) {};
        \node[seven] (lu) at (1,1) {};
        \node[seven] (u) at (2,1) {};
        \node[seven] (d) at (2,-1) {};
        \node[seven] (rd) at (3,-1) {};
        \node[seven] (r) at (3,0) {};
        \draw[magenta] (o)--(l);
        \draw[green,thick,transform canvas={yshift=3pt}] (l)--(r);
        \draw[olive,thick,transform canvas={yshift=-3pt}] (l)--(r);
        \draw[green,thick,transform canvas={xshift=3pt}] (u)--(d);
        \draw[olive,thick,transform canvas={xshift=-3pt}] (u)--(d);
        \draw[green,thick,transform canvas={xshift=3pt,yshift=3pt}] (lu)--(rd);
        \draw[olive,thick,transform canvas={xshift=-3pt,yshift=-3pt}] (lu)--(rd);
        \node at (1.6,-0.4) {\color{olive}$2$};
        \node at (2.4,0.4) {\color{green}$2$};
    \end{tikzpicture}}$};
                \node (31) at (1.5,3.5) {
    $\scalebox{1}{\begin{tikzpicture}
        \node[seven] (o) at (0,0) {};
        \node[seven] (l) at (1,0) {};
        \node[seven] (lu) at (1,1) {};
        \node[seven] (u) at (2,1) {};
        \node[seven] (d) at (2,-1) {};
        \node[seven] (rd) at (3,-1) {};
        \node[seven] (r) at (3,0) {};
        \draw[magenta] (o)--(l);
        \draw[green,transform canvas={yshift=3pt}] (l)--(r);
        \draw[olive,thick,transform canvas={yshift=-3pt}] (l)--(r);
        \draw[green,transform canvas={xshift=3pt}] (u)--(d);
        \draw[olive,thick,transform canvas={xshift=-3pt}] (u)--(d);
        \draw[green,transform canvas={xshift=3pt,yshift=3pt}] (lu)--(rd);
        \draw[olive,thick,transform canvas={xshift=-3pt,yshift=-3pt}] (lu)--(rd);
        \node at (1.6,-0.4) {\color{olive}$3$};
    \end{tikzpicture}}$};
                \node (211) at (0,7) {
    $\scalebox{1}{\begin{tikzpicture}
        \node[seven] (o) at (0,0) {};
        \node[seven] (l) at (1,0) {};
        \node[seven] (lu) at (1,1) {};
        \node[seven] (u) at (2,1) {};
        \node[seven] (d) at (2,-1) {};
        \node[seven] (rd) at (3,-1) {};
        \node[seven] (r) at (3,0) {};
        \draw[magenta] (o)--(l);
        \draw[green,transform canvas={yshift=3pt}] (l)--(r);
        \draw[green,transform canvas={yshift=1pt}] (l)--(r);
        \draw[olive,thick,transform canvas={yshift=-3pt}] (l)--(r);
        \draw[green,transform canvas={xshift=3pt}] (u)--(d);
        \draw[green,transform canvas={xshift=1pt}] (u)--(d);
        \draw[olive,thick,transform canvas={xshift=-3pt}] (u)--(d);
        \draw[green,transform canvas={xshift=3pt,yshift=3pt}] (lu)--(rd);
        \draw[green,transform canvas={xshift=1pt,yshift=1pt}] (lu)--(rd);
        \draw[olive,thick,transform canvas={xshift=-3pt,yshift=-3pt}] (lu)--(rd);
        \node at (1.6,-0.4) {\color{olive}$2$};
    \end{tikzpicture}}$};
                \node (1111) at (0,10.5) {
    $\scalebox{1}{\begin{tikzpicture}
        \node[seven] (o) at (0,0) {};
        \node[seven] (l) at (1,0) {};
        \node[seven] (lu) at (1,1) {};
        \node[seven] (u) at (2,1) {};
        \node[seven] (d) at (2,-1) {};
        \node[seven] (rd) at (3,-1) {};
        \node[seven] (r) at (3,0) {};
        \draw[magenta] (o)--(l);
        \draw[green,transform canvas={yshift=3pt}] (l)--(r);
        \draw[green,transform canvas={yshift=1pt}] (l)--(r);
        \draw[green,transform canvas={yshift=-1pt}] (l)--(r);
        \draw[green,transform canvas={yshift=-3pt}] (l)--(r);
        \draw[green,transform canvas={xshift=3pt}] (u)--(d);
        \draw[green,transform canvas={xshift=1pt}] (u)--(d);
        \draw[green,transform canvas={xshift=-1pt}] (u)--(d);
        \draw[green,transform canvas={xshift=-3pt}] (u)--(d);
        \draw[green,transform canvas={xshift=3pt,yshift=3pt}] (lu)--(rd);
        \draw[green,transform canvas={xshift=1pt,yshift=1pt}] (lu)--(rd);
        \draw[green,transform canvas={xshift=-1pt,yshift=-1pt}] (lu)--(rd);
        \draw[green,transform canvas={xshift=-3pt,yshift=-3pt}] (lu)--(rd);
    \end{tikzpicture}}$};
                \draw (4)--(31)--(211)--(1111) (4)--(22)--(211);
                \node (3) at (5.5,4.5) {
    $\scalebox{1}{\begin{tikzpicture}
        \node[seven] (o) at (0,0) {};
        \node[seven] (l) at (1,0) {};
        \node[seven] (lu) at (1,1) {};
        \node[seven] (u) at (2,1) {};
        \node[seven] (d) at (2,-1) {};
        \node[seven] (rd) at (3,-1) {};
        \node[seven] (r) at (3,0) {};
        \draw[magenta] (o)--(l);
        \draw[orange,transform canvas={yshift=3pt}] (l)--(r);
        \draw[olive, thick,transform canvas={yshift=-3pt}] (l)--(r);
        \draw[red,transform canvas={xshift=3pt}] (u)--(d);
        \draw[olive,thick,transform canvas={xshift=-3pt}] (u)--(d);
        \draw[blue,transform canvas={xshift=3pt,yshift=3pt}] (lu)--(rd);
        \draw[olive,thick,transform canvas={xshift=-3pt,yshift=-3pt}] (lu)--(rd);
        \node at (1.6,-0.4) {\color{olive}$3$};
    \end{tikzpicture}}$};
                \node (21) at (5,8) {
    $\scalebox{1}{\begin{tikzpicture}
        \node[seven] (o) at (0,0) {};
        \node[seven] (l) at (1,0) {};
        \node[seven] (lu) at (1,1) {};
        \node[seven] (u) at (2,1) {};
        \node[seven] (d) at (2,-1) {};
        \node[seven] (rd) at (3,-1) {};
        \node[seven] (r) at (3,0) {};
        \draw[magenta] (o)--(l);
        \draw[orange,transform canvas={yshift=3pt}] (l)--(r);
        \draw[green,transform canvas={yshift=1pt}] (l)--(r);
        \draw[olive, thick,transform canvas={yshift=-3pt}] (l)--(r);
        \draw[red,transform canvas={xshift=3pt}] (u)--(d);
        \draw[green,transform canvas={xshift=1pt}] (u)--(d);
        \draw[olive,thick,transform canvas={xshift=-3pt}] (u)--(d);
        \draw[blue,transform canvas={xshift=3pt,yshift=3pt}] (lu)--(rd);
        \draw[green,transform canvas={xshift=1pt,yshift=1pt}] (lu)--(rd);
        \draw[olive,thick,transform canvas={xshift=-3pt,yshift=-3pt}] (lu)--(rd);
        \node at (1.6,-0.4) {\color{olive}$2$};
    \end{tikzpicture}}$};
                \node (111) at (5,11.5) {
    $\scalebox{1}{\begin{tikzpicture}
        \node[seven] (o) at (0,0) {};
        \node[seven] (l) at (1,0) {};
        \node[seven] (lu) at (1,1) {};
        \node[seven] (u) at (2,1) {};
        \node[seven] (d) at (2,-1) {};
        \node[seven] (rd) at (3,-1) {};
        \node[seven] (r) at (3,0) {};
        \draw[magenta] (o)--(l);
        \draw[orange,transform canvas={yshift=3pt}] (l)--(r);
        \draw[green,transform canvas={yshift=1pt}] (l)--(r);
        \draw[green,transform canvas={yshift=-1pt}] (l)--(r);
        \draw[green,transform canvas={yshift=-3pt}] (l)--(r);
        \draw[red,transform canvas={xshift=3pt}] (u)--(d);
        \draw[green,transform canvas={xshift=1pt}] (u)--(d);
        \draw[green,transform canvas={xshift=-1pt}] (u)--(d);
        \draw[green,transform canvas={xshift=-3pt}] (u)--(d);
        \draw[blue,transform canvas={xshift=3pt,yshift=3pt}] (lu)--(rd);
        \draw[green,transform canvas={xshift=1pt,yshift=1pt}] (lu)--(rd);
        \draw[green,transform canvas={xshift=-1pt,yshift=-1pt}] (lu)--(rd);
        \draw[green,transform canvas={xshift=-3pt,yshift=-3pt}] (lu)--(rd);
    \end{tikzpicture}}$};
                \draw (3)--(21)--(111);
                \node (2) at (10,9) {
    $\scalebox{1}{\begin{tikzpicture}
        \node[seven] (o) at (0,0) {};
        \node[seven] (l) at (1,0) {};
        \node[seven] (lu) at (1,1) {};
        \node[seven] (u) at (2,1) {};
        \node[seven] (d) at (2,-1) {};
        \node[seven] (rd) at (3,-1) {};
        \node[seven] (r) at (3,0) {};
        \draw[magenta] (o)--(l);
        \draw[orange,transform canvas={yshift=3pt}] (l)--(r);
        \draw[orange,transform canvas={yshift=1pt}] (l)--(r);
        \draw[green,thick,transform canvas={yshift=-3pt}] (l)--(r);
        \draw[red,transform canvas={xshift=3pt}] (u)--(d);
        \draw[red,transform canvas={xshift=1pt}] (u)--(d);
        \draw[green,thick,transform canvas={xshift=-3pt}] (u)--(d);
        \draw[blue,transform canvas={xshift=3pt,yshift=3pt}] (lu)--(rd);
        \draw[blue,transform canvas={xshift=1pt,yshift=1pt}] (lu)--(rd);
        \draw[green,thick,transform canvas={xshift=-3pt,yshift=-3pt}] (lu)--(rd);
        \node at (1.6,-0.4) {\color{green}$2$};
    \end{tikzpicture}}$};
                \node (11) at (10,12.5) {
    $\scalebox{1}{\begin{tikzpicture}
        \node[seven] (o) at (0,0) {};
        \node[seven] (l) at (1,0) {};
        \node[seven] (lu) at (1,1) {};
        \node[seven] (u) at (2,1) {};
        \node[seven] (d) at (2,-1) {};
        \node[seven] (rd) at (3,-1) {};
        \node[seven] (r) at (3,0) {};
        \draw[magenta] (o)--(l);
        \draw[orange,transform canvas={yshift=3pt}] (l)--(r);
        \draw[orange,transform canvas={yshift=1pt}] (l)--(r);
        \draw[green,transform canvas={yshift=-1pt}] (l)--(r);
        \draw[green,transform canvas={yshift=-3pt}] (l)--(r);
        \draw[red,transform canvas={xshift=3pt}] (u)--(d);
        \draw[red,transform canvas={xshift=1pt}] (u)--(d);
        \draw[green,transform canvas={xshift=-1pt}] (u)--(d);
        \draw[green,transform canvas={xshift=-3pt}] (u)--(d);
        \draw[blue,transform canvas={xshift=3pt,yshift=3pt}] (lu)--(rd);
        \draw[blue,transform canvas={xshift=1pt,yshift=1pt}] (lu)--(rd);
        \draw[green,transform canvas={xshift=-1pt,yshift=-1pt}] (lu)--(rd);
        \draw[green,transform canvas={xshift=-3pt,yshift=-3pt}] (lu)--(rd);
    \end{tikzpicture}}$};
                \draw (2)--(11);
                \node (1) at (15,13.5) {
    $\scalebox{1}{\begin{tikzpicture}
        \node[seven] (o) at (0,0) {};
        \node[seven] (l) at (1,0) {};
        \node[seven] (lu) at (1,1) {};
        \node[seven] (u) at (2,1) {};
        \node[seven] (d) at (2,-1) {};
        \node[seven] (rd) at (3,-1) {};
        \node[seven] (r) at (3,0) {};
        \draw[magenta] (o)--(l);
        \draw[orange,transform canvas={yshift=3pt}] (l)--(r);
        \draw[orange,transform canvas={yshift=1pt}] (l)--(r);
        \draw[orange,transform canvas={yshift=-1pt}] (l)--(r);
        \draw[green,transform canvas={yshift=-3pt}] (l)--(r);
        \draw[red,transform canvas={xshift=3pt}] (u)--(d);
        \draw[red,transform canvas={xshift=1pt}] (u)--(d);
        \draw[red,transform canvas={xshift=-1pt}] (u)--(d);
        \draw[green,transform canvas={xshift=-3pt}] (u)--(d);
        \draw[blue,transform canvas={xshift=3pt,yshift=3pt}] (lu)--(rd);
        \draw[blue,transform canvas={xshift=1pt,yshift=1pt}] (lu)--(rd);
        \draw[blue,transform canvas={xshift=-1pt,yshift=-1pt}] (lu)--(rd);
        \draw[green,transform canvas={xshift=-3pt,yshift=-3pt}] (lu)--(rd);
    \end{tikzpicture}}$};
                \node (0) at (20,14.5) {
    $\scalebox{1}{\begin{tikzpicture}
        \node[seven] (o) at (0,0) {};
        \node[seven] (l) at (1,0) {};
        \node[seven] (lu) at (1,1) {};
        \node[seven] (u) at (2,1) {};
        \node[seven] (d) at (2,-1) {};
        \node[seven] (rd) at (3,-1) {};
        \node[seven] (r) at (3,0) {};
        \draw[magenta] (o)--(l);
        \draw[orange,transform canvas={yshift=3pt}] (l)--(r);
        \draw[orange,transform canvas={yshift=1pt}] (l)--(r);
        \draw[orange,transform canvas={yshift=-1pt}] (l)--(r);
        \draw[orange,transform canvas={yshift=-3pt}] (l)--(r);
        \draw[red,transform canvas={xshift=3pt}] (u)--(d);
        \draw[red,transform canvas={xshift=1pt}] (u)--(d);
        \draw[red,transform canvas={xshift=-1pt}] (u)--(d);
        \draw[red,transform canvas={xshift=-3pt}] (u)--(d);
        \draw[blue,transform canvas={xshift=3pt,yshift=3pt}] (lu)--(rd);
        \draw[blue,transform canvas={xshift=1pt,yshift=1pt}] (lu)--(rd);
        \draw[blue,transform canvas={xshift=-1pt,yshift=-1pt}] (lu)--(rd);
        \draw[blue,transform canvas={xshift=-3pt,yshift=-3pt}] (lu)--(rd);
    \end{tikzpicture}}$};
                \draw[red] (31)--(3) (211)--(21)--(2) (1111)--(111)--(11)--(1)--(0);
            \end{tikzpicture}
    \caption{Brane Web Hasse diagram for $4$ $\mathrm{SU}(3)$ instantons. The magnetic quivers read from each subweb decomposition following the rules in Appendix \ref{app:braneweb} match those in Figure \ref{fig:QuiverSubtractionHasse}. The coloured numbers $n$ denote the number of merged subwebs, corresponding to $\mathrm{U}(1)$ nodes of shortness $1/n$ in the magnetic quiver.}
    \label{fig:BraneWebHasse}
\end{figure}

\end{landscape}

\paragraph{Back to the brane web.}
In the bottom brane system of \eqref{eq:braneWeb2Inst} the green subwebs (making up the $\mathrm{U}(2)$ node with an adjoint in the magnetic quiver) are of self-intersection $0$. Hence we expect a further degeneration, where the two green subwebs coincide. In the magnetic quiver the coincident green subwebs become a single $\mathrm{U}(1)$ node. The magenta coloured brane still witnesses the coincident green subwebs as a pair of subwebs, while the coincident green subwebs still sees a single magenta brane. Based on this we should read a non-simply laced magnetic quiver:\footnote{We discuss the rules of reading magnetic quivers from subweb decompositions in detail in Appendix \ref{app:braneweb}.}
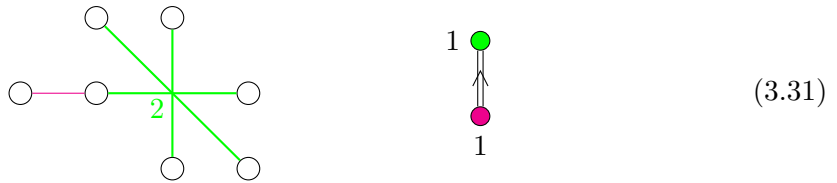
\begin{equation}
    \vcenter{\hbox{\scalebox{1}{\begin{tikzpicture}
        \node[seven] (0) at (0,0) {};
        \node[seven] (l) at (1,0) {};
        \node[seven] (lu) at (1,1) {};
        \node[seven] (u) at (2,1) {};
        \node[seven] (d) at (2,-1) {};
        \node[seven] (rd) at (3,-1) {};
        \node[seven] (r) at (3,0) {};
        \draw[magenta] (0)--(l);
        \draw[green,thick] (l)--(r);
        \draw[green,thick] (u)--(d);
        \draw[green,thick] (lu)--(rd);
        \node at (1.8,-0.2) {\color{green}$2$};
    \end{tikzpicture}}}}
    \qquad\qquad\qquad
    \vcenter{\hbox{\scalebox{1}{\begin{tikzpicture}
        \node[gaugem,label=below:{$1$}] (1) at (0,0) {};
        \node[gaugeg,label=left:{$1$}] (u) at (0,1) {};
        \draw[transform canvas={xshift=1pt}] (1)--(u);
        \draw[transform canvas={xshift=-1pt}] (1)--(u);
        \draw (0-0.1,0.5-0.1)--(0,0.5+0.1)--(0+0.1,0.5-0.1);
    \end{tikzpicture}}}}
    \label{eq:WebsLastQuiv}
\end{equation}
This magnetic quiver describes the $\mathbb{H}$ factor of the moduli space of instantons, which is exactly what we expect to find after performing all singular transitions. The green $2$ in the brane web in \eqref{eq:WebsLastQuiv} denotes, that we are dealing with two coincident subwebs. We can easily extend to any number $n$ of coincident subwebs. In this case there is a non simply laced edge of order $n$. This perfectly agrees with the quiver subtraction of the previous section.
We present the Hasse diagram of the moduli space of $4$ $\mathrm{SU}(3)$ instantons obtained from the brane web in Figure \ref{fig:BraneWebHasse}. The magnetic quivers read from each brane web decomposition are exactly those in Figure \ref{fig:QuiverSubtractionHasse}.

In Appendix \ref{app:braneweb} we present the rules to read these magnetic quivers, when coincident subwebs of self-intersection $0$ are present.

\subsection{Summary and General Case}

It is now time to summarise what we have learned about the Hasse diagram of the moduli space of instantons. We have studied the Hasse diagram of the moduli space of $4$ $\mathrm{SU}(3)$ instantons using four different approaches:
\begin{itemize}
    \item Partial Higgsing of the ADHM quiver
    \item Branes within branes
    \item Quiver subtraction of the CFHM quiver
    \item Brane web decomposition and magnetic quivers.
\end{itemize}
The Hasse diagrams produced by all four methods perfectly agree.

\paragraph{General case.}
Based on this example, we can conjecture the Hasse diagram of any $\mathcal{M}_{k,G}$:
\begin{enumerate}
    \item Start by drawing the Hasse diagrams of $\mathrm{Sym}^{k'}(\mathbb{C}^2)$ for $0\leq k'\leq k$: The leaves can be denoted by partitions of $k'$. There is an elementary transition from a partition $\lambda$ of $k'$ into $n$ parts to a partition $\mu$ of $k'$ into $n-1$ parts, if adding two parts $a,b$ in $\lambda$ yields $\mu$. If $a=b$ this is an $A_1$ transition, if $a\neq b$ this is an $m$ transition.
    \begin{equation}
        \begin{tikzpicture}
            \node (1) at (0,0) {$\lambda=[\dots,a,\dots,b,\dots]$};
            \node (2) at (0,-3) {$\mu=[\dots,a+b,\dots]$};
            \draw (1)--(2);
            \node at (1.2,-1.5) {$\left\{\begin{matrix}
                A_1 & \textnormal{if }a=b\\
                m & \textnormal{if }a\neq b
                \end{matrix}
                \right.
            $};
        \end{tikzpicture}\;.
    \end{equation}
    \item For any partition $\lambda$ of $k'<k$ there is a $g = \overline{\mathcal{O}}_{\textrm{min}} (\mathfrak{g})$ transition to the partition $\rho$ of $k'+1$ which is obtained from $\lambda$ by adding the part $1$.
    \begin{equation}
        \begin{tikzpicture}
            \node (1) at (0,0) {$\lambda=[\dots]$};
            \node (2) at (-2,-2) {$\rho=[\dots,1]$};
            \draw[red] (1)--(2);
            \node at (-0.7,-1) {$g$};
        \end{tikzpicture}\;.
    \end{equation}
\end{enumerate}

\section{Conclusion}
\label{sec:Conclusion}

The study of the Hasse diagram of the moduli space of instantons leads to the refinement of existing techniques which may be applied to a class of moduli spaces which were previously inaccessible.
The quiver subtraction algorithm for unitary quivers is extended to handle multiple subtractions of the same slice, as well as nodes with an adjoint hyper. This is accomplished through the use of decorated quivers. We in particular hope our improved quiver subtraction algorithm will allow us to study the Hasse diagram of the double affine Grassmannian \cite{Braverman:2007dvq} in future work.
When dealing with brane systems, Kraft-Procesi like transitions can be realised by coinciding branes of self-intersection zero. 

It should be noted that brute force computation of the Hasse diagram of singularities of moduli spaces of instantons is a challenge. For instance, the reduced moduli space of two $\mathrm{SU}(2)$ instantons is a complex dimension 6 affine algebraic variety which can be described by the vanishing of 15 polynomials in $\mathbb{C}^{12}$. It would be interesting to check our results in this simple case, but the computational cost seems high: the determination of the singular locus involves the computation of $\binom{15}{12}$ minors of the Jacobian matrix, each of which being a $12 \times 12$ polynomial determinant. 

Decorated quivers raise questions, beginning with their precise definition, and that of their Coulomb branches -- not only for the quivers used in the present paper, but also for multiple decorations. Coulomb branches of `conventional' quivers are always normal \cite{Braverman:2016wma}. Some of our decorated quivers, however, are deduced to have non-normal Coulomb branches. Therefore understanding decorated quivers may provide a way to construct many new interesting hyper-K\"ahler spaces. In particular, the transverse slices between arbitrary leaves in the moduli space of instantons could be described in this way. 

Finally, a precise definition of the Coulomb branch of decorated quivers should be accompanied by a method to compute the Hilbert series, generalizing the monopole formula. We plan to address these challenges in future works.

\section*{Acknowledgements}

We would like to thank Ivan Losev, Travis Schedler and Michael Shapiro for helpful comments on symmetric products. 
AB is supported by the ERC Consolidator Grant 772408-Stringlandscape, and by the LabEx ENS-ICFP: ANR-10-LABX-0010/ANR-10-IDEX-0001-02 PSL*. JFG, AH and ZZ are supported by STFC grants ST/P000762/1 and ST/T000791/1.

\clearpage

\appendix

\section{\texorpdfstring{The $G_2$ nilcone and the non-normal slice $m$}{The G2 nilcone and the non-normal slice m}} \label{app:G2}

In the study of nilpotent orbits of exceptional algebras certain elementary slices are identified to be non-normal \cite{2015arXiv150205770F}. In particular the non-normal variety $m$ \cite[§1.8.4.]{2015arXiv150205770F}, which normalises to $\mathbb{C}^2$, appears in the Hasse diagram of the nilpotent cone of every exceptional algebra, and it was in this context where $m$ was first introduced.

\paragraph{The slice m.} The slice $m$ may be characterised by its coordinate ring. 
The refined Hilbert series of $m$ is
\begin{equation}
    \mathrm{HS}(m)=\mathrm{PE}\left[[1]t\right]-[1]t = \sum_{n\not=1}^\infty[n]t^n\,,
\end{equation}
with a Plethystic Logarithm
\begin{equation}
[2]t^2+[3]t^3-t^4-\left([1]+[3]\right)t^5-([2]+[6])t^6+O(t^7) .
\end{equation}
From which we can read that the ring of functions on $m$ is generated by 7 generators transforming in [2] at degree 2 and [3] at degree 3, subject to relations at degrees 4, 5, and 6.

\paragraph{The nilpotent cone of $G_2$.} Let us study the Hasse diagram for the nilpotent cone of $G_2$ in some detail. For several slices Coulomb branch quivers can be found in the literature \cite{Hanany:2018dvd,Bourget:2020bxh}.

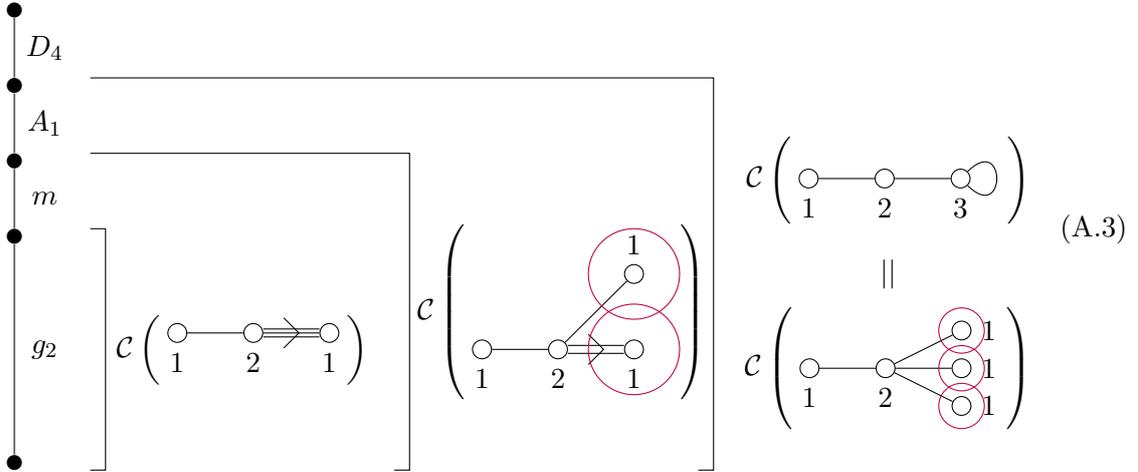
\begin{equation}
   \raisebox{-.5\height}{  \begin{tikzpicture}
        \node[hasse] (1) at (0,0) {};
        \node[hasse] (2) at (0,3) {};
        \node[hasse] (3) at (0,4) {};
        \node[hasse] (4) at (0,5) {};
        \node[hasse] (5) at (0,6) {};
        \draw (1)--(2)--(3)--(4)--(5);
        \node at (0.4,1.5) {$g_2$};
        \node at (0.4,3.5) {$m$};
        \node at (0.4,4.5) {$A_1$};
        \node at (0.4,5.5) {$D_4$};
        \draw (1,3.1)--(1.2,3.1)--(1.2,-0.1)--(1,-0.1);
        \node at (3,1.5) {$\mathcal{C}\left(\vcenter{\hbox{\scalebox{1}{\begin{tikzpicture}
            \node[gauge,label=below:{1}] (g1) at (0,0) {};
            \node[gauge,label=below:{2}] (g2) at (1,0) {};
            \node[gauge,label=below:{1}] (g3) at (2,0) {};
            \draw (g1)--(g2);
            \draw[transform canvas={yshift=-1.5pt}] (g2)--(g3);
            \draw (g2)--(g3);
            \draw[transform canvas={yshift=1.5pt}] (g2)--(g3);
            \draw (1.5-0.1,0+0.2)--(1.5+0.1,0)--(1.5-0.1,0-0.2);
        \end{tikzpicture}}}}\right)$};
        \draw (1,4.1)--(5.2,4.1)--(5.2,-0.1)--(5,-0.1);
        \node at (7.2,2) {$\mathcal{C}\left(\vcenter{\hbox{\scalebox{1}{\begin{tikzpicture}
            \node[gauge,label=below:{1}] (g1) at (0,0) {};
            \node[gauge,label=below:{2}] (g2) at (1,0) {};
            \node[gauge,label=below:{1}] (g3) at (2,0) {};
            \node[gauge,label=above:{1}] (g4) at (2,1) {};
            \draw (g1)--(g2)--(g4);
            \draw[transform canvas={yshift=-1.5pt}] (g2)--(g3);
            \draw[transform canvas={yshift=1.5pt}] (g2)--(g3);
            \draw (1.5-0.1,0+0.2)--(1.5+0.1,0)--(1.5-0.1,0-0.2);
            \draw[purple] (g3) circle (0.6cm);
            \draw[purple] (g4) circle (0.6cm);
        \end{tikzpicture}}}}\right)$};
        \draw (1,5.1)--(9.2,5.1)--(9.2,-0.1)--(9,-0.1);
        \node at (11.5,3.75) {$\mathcal{C}\left(\vcenter{\hbox{\scalebox{1}{\begin{tikzpicture}
            \node[gauge,label=below:{1}] (g1) at (0,0) {};
            \node[gauge,label=below:{2}] (g2) at (1,0) {};
            \node[gauge,label=below:{3}] (g3) at (2,0) {};
            \draw (g1)--(g2)--(g3);
            \draw (g3) to [out=315,in=45,looseness=10] (g3);
        \end{tikzpicture}}}}\right)$};
        \node at (11.5,1.25) {$\mathcal{C}\left(\vcenter{\hbox{\scalebox{1}{\begin{tikzpicture}
            \node[gauge,label=below:{1}] (g1) at (0,0) {};
            \node[gauge,label=below:{2}] (g2) at (1,0) {};
            \node[gauge,label=right:{1}] (g3) at (2,0) {};
            \node[gauge,label=right:{1}] (g3u) at (2,0.5) {};
            \node[gauge,label=right:{1}] (g3d) at (2,-0.5) {};
            \draw (g1)--(g2)--(g3) (g3u)--(g2)--(g3d);
            \draw[purple] (g3) circle (0.3cm);
            \draw[purple] (g3u) circle (0.3cm);
            \draw[purple] (g3d) circle (0.3cm);
        \end{tikzpicture}}}}\right)$};
        \node at (11.5,2.5) {$\vert\vert$};
    \end{tikzpicture}}
    \label{eq:G2Nilcone}
\end{equation}
From this diagram we can deduce the transition:

\begin{equation}
     \raisebox{-.5\height}{\begin{tikzpicture}
        \node (up) at (0,3) {$\mathcal{C}\left(\vcenter{\hbox{\scalebox{1}{\begin{tikzpicture}
            \node[gauge] (g2) at (1,0) {$\mathsf{Q}$};
            \node[gauge,label=right:{1}] (g3) at (2,0) {};
            \node[gauge,label=right:{1}] (g4) at (2,1) {};
            \draw (g2)--(g4);
            \draw[transform canvas={yshift=-1.5pt}] (g2)--(g3);
            \draw[transform canvas={yshift=1.5pt}] (g2)--(g3);
            \draw (1.5-0.1,0+0.2)--(1.5+0.1,0)--(1.5-0.1,0-0.2);
            \draw[purple] (g3) circle (0.6cm);
            \draw[purple] (g4) circle (0.6cm);
        \end{tikzpicture}}}}\right)$};
        \node (down) at (0,0) {$\mathcal{C}\left(\vcenter{\hbox{\scalebox{1}{\begin{tikzpicture}
            \node[gauge] (g2) at (1,0) {$\mathsf{Q}$};
            \node[gauge,label=right:{1}] (g3) at (2,0) {};
            \draw[transform canvas={yshift=-1.5pt}] (g2)--(g3);
            \draw (g2)--(g3);
            \draw[transform canvas={yshift=1.5pt}] (g2)--(g3);
            \draw (1.5-0.1,0+0.2)--(1.5+0.1,0)--(1.5-0.1,0-0.2);
        \end{tikzpicture}}}}\right)$};
        \draw (up)--(down);
        \node at (0.4,1.3) {$m$};
    \end{tikzpicture}}
\end{equation}
And in the particular setting $\mathsf{Q}=\vcenter{\hbox{\scalebox{1}{\begin{tikzpicture}
            \node[gauge,label=right:{$1$}] at (1,0) {};\end{tikzpicture}}}}$, since $\mathcal{C}\left(\vcenter{\hbox{\scalebox{1}{\begin{tikzpicture}
            \node[gauge,label=left:{$1$}] (g2) at (1,0) {};
            \node[gauge,label=right:{1}] (g3) at (2,0) {};
            \draw[transform canvas={yshift=-1.5pt}] (g2)--(g3);
            \draw (g2)--(g3);
            \draw[transform canvas={yshift=1.5pt}] (g2)--(g3);
            \draw (1.5-0.1,0+0.2)--(1.5+0.1,0)--(1.5-0.1,0-0.2);
        \end{tikzpicture}}}}\right)=\mathbb{H}$, we can deduce that:
        
\begin{equation}
    \mathcal{C}\left(\vcenter{\hbox{\scalebox{1}{\begin{tikzpicture}
            \node[gauge,label=left:{$1$}] (g2) at (1,0) {};
            \node[gauge,label=right:{1}] (g3) at (2,0) {};
            \node[gauge,label=right:{1}] (g4) at (2,1) {};
            \draw (g2)--(g4);
            \draw[transform canvas={yshift=-1.5pt}] (g2)--(g3);
            \draw[transform canvas={yshift=1.5pt}] (g2)--(g3);
            \draw (1.5-0.1,0+0.2)--(1.5+0.1,0)--(1.5-0.1,0-0.2);
            \draw[purple] (g3) circle (0.6cm);
            \draw[purple] (g4) circle (0.6cm);
        \end{tikzpicture}}}}\right)=m\times\mathbb{H}
        \label{eq:HSofM}
\end{equation}
At this time it is not clear how to write down the monopole formula for decorated quivers. However, in \cite{Bourget:2020bxh} the Hilbert Series of all three Coulomb branches in \eqref{eq:G2Nilcone} were computed using various techniques.

\paragraph{More about $G_2$.}

Several more slices in the nilcone of $G_2$ are known. The Higgs branch
\begin{equation}
\mathcal{H}\left(\vcenter{\hbox{\scalebox{1}{\begin{tikzpicture}
            \node[gauge,label=below:{1}] (g1) at (0,0) {};
            \node[gauge,label=below:{2}] (g2) at (1,0) {};
            \node[gauge,label=below:{3}] (g3) at (2,0) {};
            \draw (g1)--(g2)--(g3);
            \draw (g3) to [out=315,in=45,looseness=10] (g3);
        \end{tikzpicture}}}}\right)
\end{equation}
is a $3$ (quaternionic) dimensional hypersurface in $\mathbb{C}^7$, studied in \cite[Example 1.5]{2010arXiv1002.4107L}. It is identified there as the intersection of the Slodowy slice to the minimal nilpotent orbit of $G_2$ with the nilpotent cone. The $2$ (quaternionic) dimensional singular locus of this variety is given in the same paper \cite[p.\ 22]{2010arXiv1002.4107L}, let us call it $S$. The singular locus of this singular locus $S$ is expected to be $m$. The refined Hilbert series for the first two varieties can be computed from the relations of \cite{2010arXiv1002.4107L}, while the Hilbert Series of $m$ is mentioned above. It is instructive to compare the Plethystic Logarithm of the Hilbert series of the three varieties:

\begin{equation}
    \begin{tikzpicture}
        \node[hasse] (1) at (0,0) {};
        \node[hasse] (2) at (0,3) {};
        \node[hasse] (3) at (0,4) {};
        \node[hasse] (4) at (0,5) {};
        \node[hasse] (5) at (0,6) {};
        \draw (1)--(2)--(3)--(4)--(5);
        \node at (0.4,1.5) {$g_2$};
        \node at (0.4,3.5) {$m$};
        \node at (0.4,4.5) {$A_1$};
        \node at (0.4,5.5) {$D_4$};
        \draw (1,4.1)--(1.2,4.1)--(1.2,2.9)--(1,2.9);
        \node at (3.2,3.5) {$
            \begin{aligned}
                &[2]t^2+[3]t^3\\
                &-[0]t^4-([1]+[3])t^5\\
                &-([2]+[6])t^6+O(t^7)
            \end{aligned}
        $};
        \node at (3.2,2) {$=\mathrm{PL}(\mathrm{HS}(m))$};
        \draw (1,5.1)--(5.7,5.1)--(5.7,2.9)--(5.5,2.9);
        \node at (7.2,4) {$
            \begin{aligned}
                &[2]t^2+[3]t^3\\
                &-[1]t^5\\
                &-[2]t^6+O(t^8)
            \end{aligned}
        $};
        \node at (7.2,2.5) {$=\mathrm{PL}(\mathrm{HS}(S))$};
        \draw (1,6.1)--(9.2,6.1)--(9.2,2.9)--(9,2.9);
        \node at (11,4.5) {$
            \begin{aligned}
                &[2]t^2+[3]t^3\\
                &-[0]t^{12}
            \end{aligned}
        $};
        \node at (13,3) {$=\mathrm{PL}\left(\mathrm{HS}\left(\mathcal{H}\left(\vcenter{\hbox{\scalebox{1}{\begin{tikzpicture}
            \node[gauge,label=below:{1}] (g1) at (0,0) {};
            \node[gauge,label=below:{2}] (g2) at (1,0) {};
            \node[gauge,label=below:{3}] (g3) at (2,0) {};
            \draw (g1)--(g2)--(g3);
            \draw (g3) to [out=315,in=45,looseness=10] (g3);
        \end{tikzpicture}}}}\right)\right)\right)$};
    \end{tikzpicture}
    \label{eq:G2Nilcone2}
\end{equation}
We see that the generators of the three varieties are the same, but as we go down in the Hasse diagrams more relations are added.

\section{\texorpdfstring{Singularities in $\mathrm{Sym}^{k}(\mathbb{C}^2)$}{Singularities in the symmetric products of C2}}
\label{app:SYM}

In order to understand the moduli space of instantons, or rather its Hasse diagram, we must first understand symmetric products of $\mathbb{C}^2$. One way to construct this space is as the Higgs branch of an ADHM quiver:
\begin{equation}
    \mathrm{Sym}^{k}(\mathbb{C}^2)=\mathcal{H}\left(
\vcenter{\hbox{\scalebox{1}{
\begin{tikzpicture}
                \node[flavour,label=below:{$1$}] (f) at (0,0) {};
                \node[gaugeBig, label=below:{$\mathrm{U}(k)$}] (g) at (1,0) {};
                \draw (f)--(g);
                \draw (g) to [out=315,in=45,looseness=10] (g);
            \end{tikzpicture}
}}}\right)\;.
\end{equation}
This space contains a free $\mathbb{C}^2$ factor: in this appendix we focus on the space $\mathrm{Sym}^{k}(\mathbb{C}^2) / \mathbb{C}^2$. The singular part of it may alternatively be constructed as the Higgs branch of an $A_{k}$ SYM theory with $16$ supercharges. The purpose of this appendix is to study this orbifold, and in particular to compute the Hasse diagram of its singularities with the elementary transverse slices.

\subsection{Hilbert Series}

The Hilbert series for the Higgs branch is computed using the Molien formula, 
\begin{equation}
\label{HSSym}
H_{k}(t) = \frac{1}{k!} \sum\limits_{ \sigma \in S_{k}} \frac{1}{\det \left( \mathbf{1}_{2k-2} - t (\rho (\sigma) \oplus \rho (\sigma)) \right)}  \, , 
\end{equation}
where $\rho$ denotes the $k-1$ dimensional irreducible representation of $S_k$. 
Equivalently, one can use the formula for the Hilbert series $H_{k} (t)$ of $\mathrm{Sym}^{k} (X) / X$ in terms of the Hilbert series $f(t)$ of $X$, given by 
\begin{equation}
    H_{k} (t) = \frac{1}{f(t)} \frac{1}{k!} \sum\limits_{ \sigma \in S_{k}} \prod\limits_{\ell \in \lambda (\sigma)} f(t^{\ell}) \, , 
\end{equation}
where $\lambda (\sigma)$ is the partition of $k$ given by the lengths of the cycles composing the permutation $\sigma$. In the present case, the computation of the Hilbert series for $\mathrm{Sym}^{k} (\mathbb{C}^2)$ uses $f(t) = \frac{1}{(1-t)^2}$. One can also add a fugacity $u$ for the $\mathfrak{sp}(1)$ rotating the two copies of $\mathbb{C}$ in the orbifold, i.e. give charge $+1$ to the first $\rho (\sigma)$ and $-1$ to the second $\rho (\sigma))$ in (\ref{HSSym}). This yields a Hilbert series $H_{k} (t,u)$ whose $t$-expansion has coefficients which are characters of $\mathfrak{sp}(1)$, and as such can be expressed in terms of the characters $[n]$ of the $(n+1)$-dimensional irreducible representation of $\mathfrak{sp}(1)$. 

Explicit computation of $H_{k}(t)$ for small values of $k$ lead to the following observation, which we conjecture to be valid for all $k$: 
\begin{equation}
\label{plogGenericN}
    \mathrm{PL} (H_{k}(t , u)) = \left(  \sum\limits_{j=2}^{k} [j] t^{j} \right) - [k-2]t^{k+2} + O (t^{k+3}) \, . 
\end{equation}

\paragraph{Example $k=4$}
For instance, using either method one finds 
\begin{equation}
H_4(t) = \frac{1+t^2+2 t^3+4 t^4+2 t^5+4 t^6+2 t^7+4 t^8+2 t^9+t^{10}+t^{12}}{(1-t^2)^2(1-t^3)^2(1-t^4)^2} \, . 
\end{equation}
The plethystic logarithm of this Hilbert series evaluates to $3 t^2+4 t^3+5 t^4-3 t^6-6 t^7-6 t^8+O(t^9)$. One can refine this computation using a fugacity $u$ for the $\mathfrak{sp}(1)$ that exchanges the two $\mathbb{C}^3$ factors. The Hilbert series becomes 
\begin{equation}
\frac{1+t^2+[1] t^3+([2]+[0]) t^4+[1] t^5+([2]+[0])  t^6+[1] t^7+([2]+[0])  t^8+[1] t^9+t^{10}+t^{12}}{(1-(tu)^2)(1-(t/u)^2)(1-(tu)^3)(1-(t/u)^3)(1-(tu)^4)(1-(t/u)^4)} 
\end{equation}
and the plog is 
\begin{equation}
[2] t^2 + [3]t^3 + [4]t^4 - [2]t^6 -  ([3]+[1])t^7 - ([4] + [0]) t^8 + O(t^9)
\end{equation}
where $[n]$ is the character of the $(n+1)$-dimensional irrep of $\mathfrak{sp}(1)$.

\subsection{Invariants}

Let us call $(x_i ,  y_i)_{i=1 , \dots , k-1}$ the standard coordinates on $\mathbb{C}^{k-1} \oplus \mathbb{C}^{k-1}$. 
We use the $\mathfrak{sl}_2$ algebra spanned by\footnote{These satisfy the standard $[\mathbf{e} ,\mathbf{f}]= \mathbf{h}$, $[\mathbf{h},\mathbf{e} ]=2\mathbf{e} $ and $[\mathbf{h},\mathbf{f} ]=-2\mathbf{f}$. }
\begin{eqnarray}
\mathbf{e} &=&  \sum\limits_{i=1}^{k-1} y_i \frac{\partial}{\partial x_i} \\
\mathbf{f} &=&  \sum\limits_{i=1}^{k-1} x_i \frac{\partial}{\partial y_i} \\
\mathbf{h} &=&  \sum\limits_{i=1}^{k-1} y_i \frac{\partial}{\partial y_i} - x_i \frac{\partial}{\partial x_i} 
\end{eqnarray}
We generate invariants in the following way. For $j=1, 3/2, \dots , \frac{k}{2}$ and $m = -j , -j+1 , \dots , j$, let 
\begin{equation}
\label{invariants}
V_{-j+m}^j = \frac{\mathbf{e}^m}{m!}  \langle x_1^{2j} \rangle_{S_{k}} 
\end{equation}
where $\langle \cdot \rangle_{S_{k}}$ denotes the average over the group. 
The plethystic logarithm (\ref{plogGenericN}) computed above suggests this is a complete set of $\frac12 (k-1)(k+4)$ invariants, and this can be checked at small $k$ by an explicit computation with \texttt{Macaulay2} \cite{Macaulay2}. 

\subsection{Hasse Diagram}

The fixed loci $ \{ (x_i ,  y_i)_{i=1 , \dots , k-1} \}$ in $\mathbb{C}^{k-1} \oplus \mathbb{C}^{k-1}$ under the $S_{k}$ descend in the orbifold to loci which are in one-to-one correspondence with conjugacy classes of $S_{k}$, i.e. integer partitions of $k$. We call $V(x_i ,  y_i)$ the vector of $\frac12 (k-1)(k+4)$ invariants defined in (\ref{invariants}). 

Let $\lambda = \{ \lambda_j \}_{j = 1 , \dots , r}$, with $ \lambda_1 \geq \dots \geq \lambda_r$ be an integer partition of $k$, and let $s_j = \lambda_1 + \dots + \lambda_j$ be the partial sums. We call $V_{\lambda}(x_{1} , \dots , x_{r-1} ,  y_{1} , \dots , y_{r-1})$ the vector $V(x_1 , \dots , x_k ,  y_1 , \dots , y_k)$ after the substitutions $x_{l} \rightarrow x_j$ and $y_{l} \rightarrow y_j$ for all $s_{j-1} < l \leq s_j$. Note that $V_{[1^{k}]} = V$. The vector $V_{\lambda}(x_{1} , \dots , x_{r-1} ,  y_{1} , \dots , y_{r-1})$ parametrizes the $2(r-1)$ complex dimensional variety $\mathcal{L}_{\lambda}$ in the orbifold associated to the partition $\lambda$. 

The inclusion relation between the varieties $\mathcal{L}_{\lambda}$ defines a partial order. The Hasse diagram of this partial order is the singularity Hasse diagram we are after. There is an elementary slice transverse to $\mathcal{L}_{\lambda '}$ in $\mathcal{L}_{\lambda}$ if $\lambda '$ can be obtained from $\lambda$ by replacing two elements of $\lambda$ by their sum. For simplicity, we denote this process as\footnote{We do not assume here that $\lambda_1$ and $\lambda_2$ are the largest elements of $\lambda$, nor that they are adjacent. We place them first for ease of notation. } $(\lambda_1 , \lambda_2 , \dots) \rightarrow (\lambda_1 + \lambda_2 , \dots)$. In order to compute the geometry of the transverse slice, one then uses an adapted parametrization for the $x_1, y_1, x_2, y_2$ coordinates in $V_{\lambda}$:  
\begin{equation}
\label{changeVar}
    x_1 = a (1 + \lambda_2 u), x_2 = a (1 - \lambda_1 u), 
    y_1 = b (1 + \lambda_2 v), y_2 = b (1 - \lambda_1 v). 
\end{equation}
In these coordinates, $\mathcal{L}_{\lambda '}$ is given by the equations $u=v=0$ in $\mathcal{L}_{\lambda }$. The two-complex dimensional transverse slice coordinate ring is generated by the components of 
\begin{equation}
\label{eqTransverseSlice}
    V_{\lambda} - V_{\lambda}|_{u=v=0}
\end{equation}
seen as polynomials in $u$ and $v$ with coefficients in polynomials in the other variables. 
We illustrate these computations for $k=3,4$ in the next subsections (the case $k=2$ is known).

\subsubsection*{\texorpdfstring{Case $k=3$}{Case k=3}}

The vectors $V_{\lambda}$ are 
\begin{equation}
V_{[1^3]} = \left(
\begin{array}{c}
 \frac{2}{3} \left(x_1^2+x_1 x_2+x_2^2\right) \\
 \frac{2}{3} (2 x_1 y_1+x_1 y_2+x_2 y_1+2 x_2 y_2) \\
 \frac{2}{3} \left(y_1^2+y_1 y_2+y_2^2\right) \\
 -x_1 x_2 (x_1+x_2) \\
 x_1^2 (-y_2)-2 x_1 x_2 y_1-2 x_1 x_2 y_2-x_2^2
   y_1 \\
 -2 x_1 y_1 y_2-x_1 y_2^2-x_2 y_1^2-2 x_2 y_1
   y_2 \\
 -y_1 y_2 (y_1+y_2) \\
\end{array}
\right) \, , \quad V_{[2,1]} = \left(
\begin{array}{c}
 2 x_1^2 \\
 4 x_1 y_1 \\
 2 y_1^2 \\
 -2 x_1^3 \\
 -6 x_1^2 y_1 \\
 -6 x_1 y_1^2 \\
 -2 y_1^3 \\
\end{array}
\right) \, , \quad V_{[3]} = \left(
\begin{array}{c}
0 \\
0\\
0\\
0 \\
0 \\
0 \\
0 \\
\end{array}
\right) 
\end{equation}

The transition $[1^3] \rightarrow [2,1]$ is obtained by performing the change of variables (\ref{changeVar}) with $\lambda_1 = \lambda_2 = 1$ in $V_{[1^3]}$ and evaluating (\ref{eqTransverseSlice}), giving 
\begin{equation}
    V_{[1^3]} (a(1+u), a(1-u), b(1+v), b(1-v)) -  V_{[1^3]} (a,a,b,b) =  \left(
\begin{array}{c}
 \frac{2 a^2 u^2}{3} \\
 \frac{4}{3} a b u v \\
 \frac{2 b^2 v^2}{3} \\
 2 a^3 u^2 \\
 2 a^2 b u^2+4 a^2 b u v \\
 4 a b^2 u v+2 a b^2 v^2 \\
 2 b^3 v^2 \\
\end{array}
\right) \, . 
\end{equation}
All the polynomials in this vector lie in $\mathbb{C}[a,b][u^2 , uv , v^2]$, which identifies the transverse slice as an $a_1$. The transverse slice $[2,1] \rightarrow [3]$ is identified immediately, as the coordinate ring is $\mathbb{C}[u^2 , uv , v^2 , u^3 , u^2 v , uv^2 , v^3]$: this is the slice $m$. 
Therefore the Hasse diagram is 
\begin{equation}
 \begin{tikzpicture}
                \node[hasse] (1) at (0,0) {};
                \node[hasse] (2) at (0,1) {};
                \node[hasse] (3) at (0,2) {};
                \draw (1)--(2)--(3);
                \node at (-0.4,0.5) {$m$};
                \node at (-0.4,1.5) {$a_1$};
        \end{tikzpicture}
\end{equation}

\subsubsection*{\texorpdfstring{Case $k=4$}{Case k=4}}

For $k=4$ we have 12 invariants. The expression (\ref{invariants}) is readily evaluated, but we do not reproduce it here as it is voluminous. The other $V_{\lambda}$ are: 
\begin{equation}
    V_{[2,1^2]} = \left(
\begin{array}{c}
 \frac{1}{2} \left(3 x_1^2+2 x_1
   x_2+x_2^2\right) \\
 x_1 (3 y_1+y_2)+x_2 (y_1+y_2) \\
 \frac{1}{2} \left(3 y_1^2+2 y_1
   y_2+y_2^2\right) \\
 -\frac{3}{2} x_1 (x_1+x_2)^2 \\
 -\frac{3}{2} (x_1+x_2) (3 x_1 y_1+2 x_1
   y_2+x_2 y_1) \\
 -\frac{3}{2} (y_1+y_2) (x_1 (3 y_1+y_2)+2
   x_2 y_1) \\
 -\frac{3}{2} y_1 (y_1+y_2)^2 \\
 \frac{1}{2} \left(9 x_1^4+16 x_1^3 x_2+12 x_1^2
   x_2^2+4 x_1 x_2^3+x_2^4\right) \\
 2 \left(x_1^3 (9 y_1+4 y_2)+6 x_1^2 x_2
   (2 y_1+y_2)+3 x_1 x_2^2 (2
   y_1+y_2)+x_2^3 (y_1+y_2)\right) \\
 3 \left(x_1^2 \left(9 y_1^2+8 y_1 y_2+2
   y_2^2\right)+2 x_1 x_2 (2
   y_1+y_2)^2+x_2^2 \left(2 y_1^2+2 y_1
   y_2+y_2^2\right)\right) \\
 2 \left(x_1 \left(9 y_1^3+12 y_1^2 y_2+6
   y_1 y_2^2+y_2^3\right)+x_2 \left(4
   y_1^3+6 y_1^2 y_2+3 y_1
   y_2^2+y_2^3\right)\right) \\
 \frac{1}{2} \left(9 y_1^4+16 y_1^3 y_2+12 y_1^2
   y_2^2+4 y_1 y_2^3+y_2^4\right) \\
\end{array}
\right)
\end{equation}
\begin{equation}
     V_{[2^2]} =\left(
\begin{array}{c}
 x_1^2 \\
 2 x_1 y_1 \\
 y_1^2 \\
 0 \\
 0 \\
 0 \\
 0 \\
 x_1^4 \\
 4 x_1^3 y_1 \\
 6 x_1^2 y_1^2 \\
 4 x_1 y_1^3 \\
 y_1^4 \\
\end{array}
\right) \, , \quad V_{[3,1]} = \left(
\begin{array}{c}
 3 x_1^2 \\
 6 x_1 y_1 \\
 3 y_1^2 \\
 -6 x_1^3 \\
 -18 x_1^2 y_1 \\
 -18 x_1 y_1^2 \\
 -6 y_1^3 \\
 21 x_1^4 \\
 84 x_1^3 y_1 \\
 126 x_1^2 y_1^2 \\
 84 x_1 y_1^3 \\
 21 y_1^4 \\
\end{array}
\right) \, , \quad V_{[4]} =\left(
\begin{array}{c}
 0 \\
 0 \\
 0 \\
 0 \\
 0 \\
 0 \\
 0 \\
 0 \\
 0 \\
 0 \\
 0 \\
 0 \\
\end{array}
\right) \, . 
\end{equation}

The only two non-trivial transverse slices to compute are $[2,1^2] \rightarrow [2^2]$ and $[2,1^2] \rightarrow [3,1]$. Using (\ref{eqTransverseSlice}), one computes $V_{[2,1^2]} (-a,a(1+u),-b,b(1+v)) - V_{[2,1^2]} (-a,a,-b,b)$, which gives 
\begin{equation}
 \left(
\begin{array}{c}
 \frac{a^2 u^2}{2} \\
 a b u v \\
 \frac{b^2 v^2}{2} \\
 \frac{3 a^3 u^2}{2} \\
 \frac{3}{2} a^2 b u^2+3 a^2 b u v \\
 3 a b^2 u v+\frac{3}{2} a b^2 v^2 \\
 \frac{3 b^3 v^2}{2} \\
 \frac{a^4 u^4}{2}+3 a^4 u^2 \\
 2 a^3 b u^3 v+6 a^3 b u^2+6 a^3 b u v \\
 3 a^2 b^2 u^2 v^2+3 a^2 b^2 u^2+12 a^2 b^2 u v+3 a^2 b^2 v^2 \\
 2 a b^3 u v^3+6 a b^3 u v+6 a b^3 v^2 \\
 \frac{b^4 v^4}{2}+3 b^4 v^2 \\
\end{array}
\right) \, 
\end{equation}
and $V_{[2,1^2]} (a(1+u),a(1-2u),b(1+v),b(1-2v)) - V_{[2,1^2]} (a,a,b,b)$, which gives 
\begin{equation}
     \left(
\begin{array}{c}
 \frac{3 a^2 u^2}{2} \\
 3 a b u v \\
 \frac{3 b^2 v^2}{2} \\
 \frac{9 a^3 u^2}{2}-\frac{3 a^3 u^3}{2} \\
 -\frac{9}{2} a^2 b u^2 v+\frac{9}{2} a^2 b u^2+9 a^2 b u v \\
 -\frac{9}{2} a b^2 u v^2+9 a b^2 u v+\frac{9}{2} a b^2 v^2 \\
 \frac{9 b^3 v^2}{2}-\frac{3 b^3 v^3}{2} \\
 \frac{9 a^4 u^4}{2}-6 a^4 u^3+9 a^4 u^2 \\
 18 a^3 b u^3 v-6 a^3 b u^3-18 a^3 b u^2 v+18 a^3 b u^2+18 a^3 b u v \\
 27 a^2 b^2 u^2 v^2-18 a^2 b^2 u^2 v+9 a^2 b^2 u^2-18 a^2 b^2 u v^2+36
   a^2 b^2 u v+9 a^2 b^2 v^2 \\
 18 a b^3 u v^3-18 a b^3 u v^2+18 a b^3 u v-6 a b^3 v^3+18 a b^3 v^2 \\
 \frac{9 b^4 v^4}{2}-6 b^4 v^3+9 b^4 v^2 \\
\end{array}
\right) \, . 
\end{equation}
This identifies the first slice as $a_1$ and the second slice as $m$. 
This completes the Hasse diagram: 
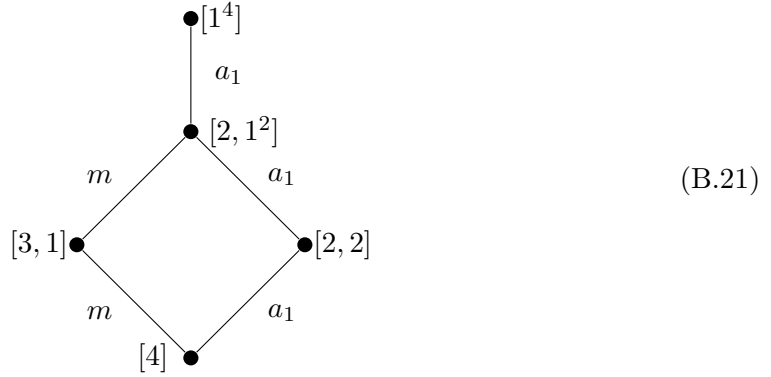
\begin{equation}
 \raisebox{-.5\height}{ \begin{tikzpicture}
                \node[hasse] (1) at (0,0) {};
                \node[hasse] (2) at (1.5,1.5) {};
                \node[hasse] (3) at (-1.5,1.5) {};
                \node[hasse] (4) at (0,3) {};
                \node[hasse] (5) at (0,4.5) {};
                \draw (4)--(2)--(1)--(3)--(4)--(5);
                \node at (-1.2,0.6) {$m$};
                \node at (1.2,2.4) {$a_1$};
                \node at (1.2,0.6) {$a_1$};
                \node at (-1.2,2.4) {$m$};
                \node at (.5,3.75) {$a_1$};
                \node at (.4,4.5) {$[1^4]$};
                \node at (.7,3) {$[2,1^2]$};
                \node at (-2,1.5) {$[3,1]$};
                \node at (2,1.5) {$[2,2]$};
                \node at (-.5,0) {$[4]$};
        \end{tikzpicture}}
\end{equation}

\subsubsection*{\texorpdfstring{General $k$}{General k}}

Based on the results of these computations, one is lead to conjecture that the transition transverse slice $\mathcal{L}_{\lambda} \rightarrow \mathcal{L}_{\lambda '}$ obtained by adding two elements of $\lambda$ is $a_1$ if the two elements are the same, and $m$ otherwise. Assuming this, one can straightforwardly draw the Hasse diagrams for $k=5$ and $k=6$: 
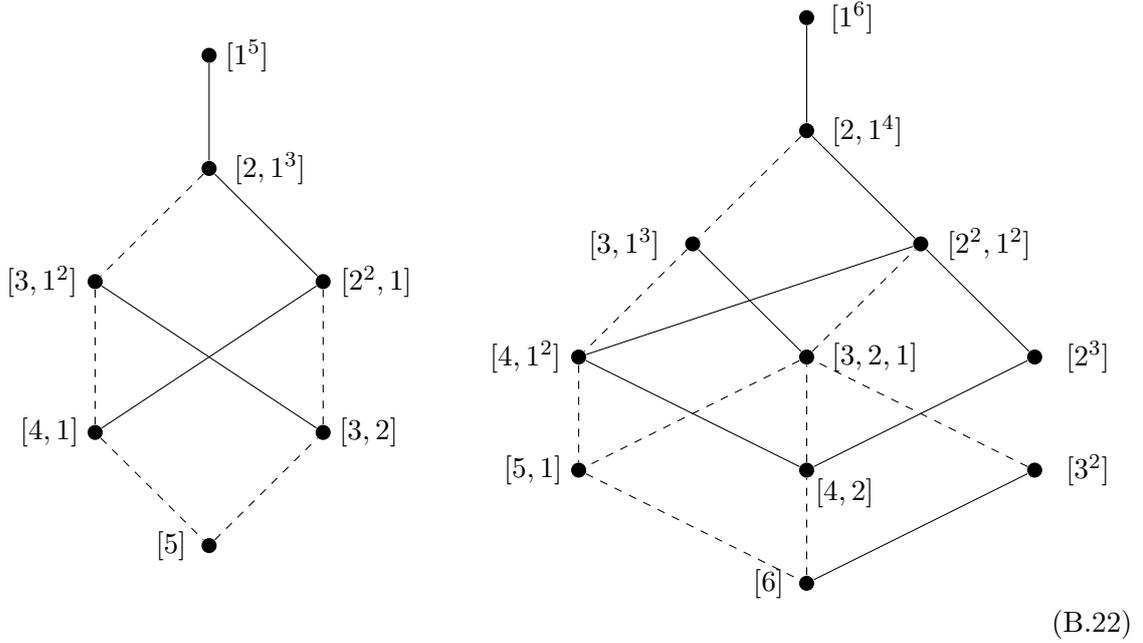
\begin{equation}
 \raisebox{-.5\height}{\begin{tikzpicture}
                \node[hasse] (1) at (0,0) {};
                \node[hasse] (2) at (-1.5,1.5) {};
                \node[hasse] (3) at (1.5,1.5) {};
                \node[hasse] (4) at (-1.5,3.5) {};
                \node[hasse] (5) at (1.5,3.5) {};
                \node[hasse] (6) at (0,5) {};
                \node[hasse] (7) at (0,6.5) {};
                \draw (7)--(6)--(5)--(2) (3)--(4);
                \draw[dashed] (6)--(4)--(2)--(1)--(3)--(5);
                \node at (.5,6.5) {$[1^5]$};
                \node at (.8,5) {$[2,1^3]$};
                \node at (-2.2,3.5) {$[3,1^2]$};
                \node at (-2.1,1.5) {$[4,1]$};
                \node at (2.1,1.5) {$[3,2]$};
                \node at (2.2,3.5) {$[2^2,1]$};
                \node at (-.5,0) {$[5]$};
        \end{tikzpicture}}
        \qquad 
          \raisebox{-.5\height}{\begin{tikzpicture}
                \node[hasse] (1) at (0,0) {};
                \node[hasse] (2) at (-3,1.5) {};
                \node[hasse] (3) at (0,1.5) {};
                \node[hasse] (4) at (-3,3) {};
                \node[hasse] (5) at (0,3) {};
                \node[hasse] (6) at (3,3) {};
                \node[hasse] (7) at (-1.5,4.5) {};
                \node[hasse] (8) at (1.5,4.5) {};
                \node[hasse] (9) at (0,6) {};
                \node[hasse] (10) at (0,7.5) {};
                \node[hasse] (11) at (3,1.5) {};
                \draw[dashed] (9)--(7)--(4)--(2)--(1)--(3)--(5)--(2) (11)--(5)--(8);
                \draw (10)--(9)--(8)--(6)--(3)--(4)--(8) (7)--(5) (1)--(11);
                \node at (.6,7.5) {$[1^6]$};
                \node at (.8,6) {$[2,1^4]$};
                \node at (-2.4,4.5) {$[3,1^3]$};
                \node at (2.4,4.5) {$[2^2,1^2]$};
                \node at (-3.7,3) {$[4,1^2]$};
                \node at (0.9,3) {$[3,2,1]$};
                \node at (3.7,3) {$[2^3]$};
                \node at (-3.6,1.5) {$[5,1]$};
                \node at (.5,1.2) {$[4,2]$};
                \node at (3.7,1.5) {$[3^2]$};
                \node at (-.5,0) {$[6]$};
        \end{tikzpicture}}
\end{equation}
where a full line represents an $a_1$ transition and a dashed line represents an $m$ transition.

\section{Magnetic quivers from brane webs with coincident branes}\label{app:braneweb}

In this section we extend the rules of \cite{Cabrera:2018jxt} and \cite[Appendix B]{Bourget:2020mez} to read the magnetic quiver associated to a brane web decomposition, in order to deal with coincident subwebs of self intersection $0$, as well as decoration. We do not consider the presence of orientifold planes.

\paragraph{Decomposition.}
Consider a brane web $W$, which is decomposed into $n$ stacks of subwebs:
\begin{equation}
W=\bigcup_{i=1}^n m_i W_i\,,
\end{equation}
where $m_i\in\mathbb{N}\backslash\{0\}$.

\paragraph{Intersection Number.}
The intersection number between two subwebs $W_i$ and $W_j$ is defined as:
\begin{equation}
    \label{eq:IntersectionNumber}
    W_i\cdot W_j=\mathrm{SI}_{i,j} + \mathrm{X}_{i,j} - \mathrm{Y}_{i,j}\,,
\end{equation}
where:
\begin{itemize}
    \item $\mathrm{SI}_{i,j}$ is the stable intersection between the tropical curves of $W_i$ and $W_j$;
    \item $\mathrm{X}_{i,j}$ is the number of pairs of five-branes (one from $W_i$ and one from $W_j$) ending on the same seven-brane from opposite sides;
    \item $\mathrm{Y}_{i,j}$ is the number of pairs of five-branes (one from $W_i$ and one from $W_j$) ending on the same seven-brane from the same side. 
\end{itemize}

\paragraph{Conditions on Decomposition.}
We require the following from our subweb decomposition:
\begin{enumerate}
    \item $W_i\cdot W_i\in\{-2,0\}$.\footnote{In general it would be fine to consider decompositions with $W_i\cdot W_i>0$. However we have little experience with them, and therefore do not wish to make any general statements.}
    \item $W_i\cdot W_j\geq0$ $\forall\,i\neq j$.
\end{enumerate}
Note that we do \emph{not} require that the $W_i$ cannot be further decomposed, in other words, the decomposition need not be maximal.

\paragraph{Elemental Subwebs and Coincidence.}
Take a decomposition of a brane web $W$ into stacks of subwebs $m_iW_i$, obeying the conditions above. Let $W_i=k^e_iW^e_i$ for some $k^e_i\in\mathbb{N}\backslash\{0\}$, s.t.\ there is no $k'_i>1$ s.t.\ $W^e_i=k'_iW'_i$. We call $W^e_i$ the \emph{elemental} of $W_i$, and $k^e_i$ the \emph{coincidence} of $W_i$. 

Note that any subweb $W_i$, with $W_i\cdot W_i=-2$, is its own elemental. In other words, we have $W_i=W^e_i$ and $k^e_i=1$. 

A crucial difference here to the decompositions mentioned in \cite[Appendix B]{Bourget:2020mez} is the possibility of $k^e_i>1$ when the self-intersection is $0$. It is important to keep track of the coincidence of brane webs with self-intersection 0, as it corresponds to singularities in the moduli space. This is the brane web analogue of the phenomenon illustrated for D3 branes in Figure \ref{fig:TwoSystems}. As is discussed below, this leads to short nodes in the magnetic quiver.

\paragraph{Magnetic Quiver.}
The decomposition of a brane web defines a magnetic quiver in terms of an $n$-vector $v$ of gauge ranks, an $n\times n$ adjacency matrix $A$, and a decoration. We have
\begin{equation}
    v_i=m_i
\end{equation}
and
\begin{equation}
    A_{ij}=\frac{W_i\cdot W_j}{k^e_j}\;.
\end{equation}
Note that the adjacency matrix does not have to be symmetric, depending on the coincidences $k^e_j$. If $A_{ij} = A_{ji}$, there are $A_{ij}$ links between nodes $i$ and $j$; if $A_{ij} > A_{ji}$ there is a non simply laced edge of order $\frac{A_{ij}}{A_{ji}}$ from node $j$ (long) to node $i$ (short). In addition, the node $i$ carries $\frac{1}{2} (2 + A_{ii})$ adjoint loops. 

In the magnetic quiver, nodes $n_i$ ($i\in I\subset\{1,\dots,n\}$), corresponding to subwebs $W_i$, are decorated in the same colour as a node $n_j$ ($j \notin I$), corresponding to the subweb $W_j$, if the elemental $W^e_j$ can be constructed by combining the elementals $W^e_i$, i.e. if
\begin{equation}
    W^e_j=\bigcup_{i\in I} W^e_i\,.
\end{equation}

\paragraph{Example:} Let us consider the brane web decomposition
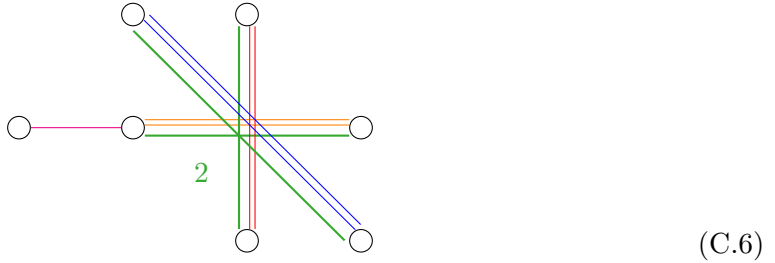
\begin{equation}
    \begin{tikzpicture}[scale=1.5]
        \node[seven] (o) at (0,0) {};
        \node[seven] (l) at (1,0) {};
        \node[seven] (lu) at (1,1) {};
        \node[seven] (u) at (2,1) {};
        \node[seven] (d) at (2,-1) {};
        \node[seven] (rd) at (3,-1) {};
        \node[seven] (r) at (3,0) {};
        \draw[magenta] (o)--(l);
        \draw[orange,transform canvas={yshift=3pt}] (l)--(r);
        \draw[orange,transform canvas={yshift=1pt}] (l)--(r);
        \draw[goodgreen,thick,transform canvas={yshift=-3pt}] (l)--(r);
        \draw[red,transform canvas={xshift=3pt}] (u)--(d);
        \draw[red,transform canvas={xshift=1pt}] (u)--(d);
        \draw[goodgreen,thick,transform canvas={xshift=-3pt}] (u)--(d);
        \draw[blue,transform canvas={xshift=3pt,yshift=3pt}] (lu)--(rd);
        \draw[blue,transform canvas={xshift=1pt,yshift=1pt}] (lu)--(rd);
        \draw[goodgreen,thick,transform canvas={xshift=-3pt,yshift=-3pt}] (lu)--(rd);
        \node at (1.6,-0.4) {\color{goodgreen}$2$};
    \end{tikzpicture}
\end{equation}
We identify the web decomposition using the colouring of the subwebs:
\begin{equation}
    W={\color{goodgreen}W_1}\cup{\color{magenta}W_2}\cup2{\color{orange}W_3}\cup2{\color{red}W_4}\cup2{\color{blue}W_5}\,.
\end{equation}
Note that ${\color{goodgreen}W_1}$ consists of {\color{goodgreen}$2$} coincident subwebs, hence $k^e_1=2$. For $i\neq1$ we have $k^e_i=1$. We find:
\begin{equation}
    v=\begin{pmatrix}
    1\\
    1\\
    2\\
    2\\
    2
    \end{pmatrix}
\end{equation}
and
\begin{equation}
    A=\begin{pmatrix}
    0 & 2 & 0 & 0 & 0 \\
    1 & -2 & 1 & 0 & 0 \\
    0 & 1 & -2 & 1 & 1 \\
    0 & 0 & 1 & -2 & 1 \\
    0 & 0 & 1 & 1 & -2
    \end{pmatrix}\,.
\end{equation}
The nodes corresponding to $W_3$, $W_4$, and $W_5$ are decorated in the same colour as $W_1$, since
\begin{equation}
    W^e_1=W_3\cup W_4\cup W_5\,.
\end{equation}
This leads to the magnetic quiver:
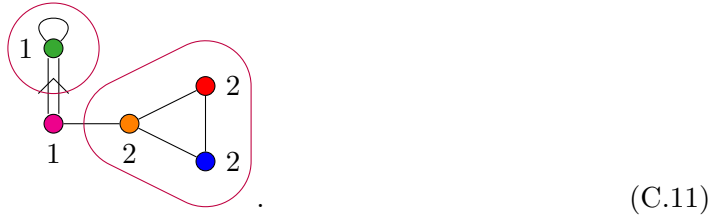
\begin{equation}
    \begin{tikzpicture}
        \node[gaugem,label=below:{$1$}] (a) at (-1,0) {};
        \node[gaugeo,label=below:{$2$}] (b) at (0,0) {};
        \node[gauger,label=right:{$2$}] (c) at (1,0.5) {};
        \node[gaugeb,label=right:{$2$}] (d) at (1,-0.5) {};
        \node[gaugegoodgreen,label=left:{$1$}] (u) at (-1,1) {};
        \draw (a)--(b)--(c)--(d)--(b);
        \draw[transform canvas={xshift=2pt}] (a)--(u);
        \draw[transform canvas={xshift=-2pt}] (a)--(u);
        \draw (-1-0.2,0.5-0.1)--(-1,0.5+0.1)--(-1+0.2,0.5-0.1);
        \draw[purple] (u) circle (0.6cm);
        \draw[purple] \convexpath{b,c,d}{0.6cm};
        \draw (u) to [out=45,in=135,looseness=8] (u);
    \end{tikzpicture}\,.
\end{equation}
The adjoint on the $\mathrm{U}(1)$ node (due to $A_{11}=0$) does not affect the Coulomb branch, and we only depict it here for clarity.

\bibliographystyle{JHEP}
\bibliography{bibli.bib}

\end{document}